\documentclass[a4paper,12pt]{scrartcl} 
\title{Hierarchical clustering: 
       visualization, feature importance and 
       model selection}

\author{Luben M. C. Cabezas,
Rafael Izbicki,
Rafael B. Stern}

\usepackage[T1]{fontenc}
\usepackage{nameref}
\usepackage[colorlinks=true, urlcolor=blue, citecolor=blue, linkcolor=blue]{hyperref}

\usepackage[textfont=bf]{caption}  % <--- new
\usepackage{subcaption}
\usepackage{siunitx}               % <--- new
\usepackage{tikz}
\usepackage{color}
\usepackage{soul}

\usepackage{amsfonts, amsmath, amssymb, amsthm, thmtools}
\usepackage{mathrsfs}
\usepackage{cleveref}

\usepackage{algorithm, algpseudocode}

\renewcommand{\algorithmicrequire}{\textbf{\small Input:}}
\renewcommand{\algorithmicensure}{\textbf{\small Output:}}

\crefname{section}{section}{sections}
\Crefname{section}{Section}{Sections}
\crefname{table}{table}{tables}
\Crefname{table}{Table}{Tables}

\usepackage{color, enumitem, float, fourier, layout, multirow, setspace, verbatim}
\setlist[enumerate]{leftmargin=*}
\usepackage[colorinlistoftodos]{todonotes}
\usepackage{listings}
\usepackage{caption,subcaption}
\usepackage{epigraph}
\usepackage{ulem}
\usepackage{makecell}

\usepackage[round]{natbib}

\usepackage{graphicx}
\graphicspath{{figures/}{../figures/}}
\def\I{{\mathbb{I}}}

\def\T{{\mathcal T}}

\usepackage{adjustbox}
\usepackage{multirow}

\usepackage{makecell}

\definecolor{darkgreen}{rgb}{0.3, 0.5, 0.0}

\newcommand{\add}[1]{#1}

\usepackage{mathtools}
\usepackage{ifthen}

\renewcommand{\emph}[1]{\textit{#1}}

\begin{document}

\maketitle

\begin{abstract}
We propose methods for the analysis of
hierarchical clustering that 
fully use the multi-resolution structure
provided by a dendrogram.
Specifically, we propose 
a loss for choosing between clustering methods,
a feature importance score and
a graphical tool for visualizing
the segmentation of features in a dendrogram.
Current approaches to these tasks 
lead to loss of information since they 
require the user to generate a single partition of 
the instances by 
cutting the dendrogram at a specified level.
Our proposed methods, instead,
use the full structure of the dendrogram.
The key insight behind the proposed methods is 
to view a dendrogram as a phylogeny.
This analogy permits the assignment 
of a feature value to each internal node of a tree
through \add{an evolutionary model.}
Real and simulated datasets provide evidence that
our proposed framework has desirable outcomes \add{and gives more insights than state-of-art approaches.}
We provide an R package
that implements our methods.
\end{abstract}

%\begin{keywords}

%\end{keywords}

\section{Introduction}
\label{introducao}

Clustering methods have the goal of
grouping similar sample points.
They are used in applications such as 
pattern recognition  \citep{Chen2014}, 
image analysis \citep{Rocha2009}, 
bioinformatics \citep{Datta2003}, and
information retrieval \citep{jardine1971}. 

Clustering techniques can be divided into 
two categories: partition-based and hierarchical.
An example of partition-based clustering is
K-means \citep{Macqueen1967}, which creates 
$K$ clusters based on 
a Voronoi partition of the feature space.
Similarly, mode-based clustering \citep{Fukunaga1975} 
creates a partition by assigning
each observation to a mode of a density estimate.
On the other hand, hierarchical clustering
yields a tree-based representation of the objects (dendrogram),
such as in agglomerative clustering \citep{ward1963hierarchical}, DIANA \citep{Rousseeuw1990} and
partial least squares methods \citep{Liu2006}.
This paper focuses solely on hierarchical clustering.

\begin{figure}[h]
 \centering
 \includegraphics[scale = 0.75]{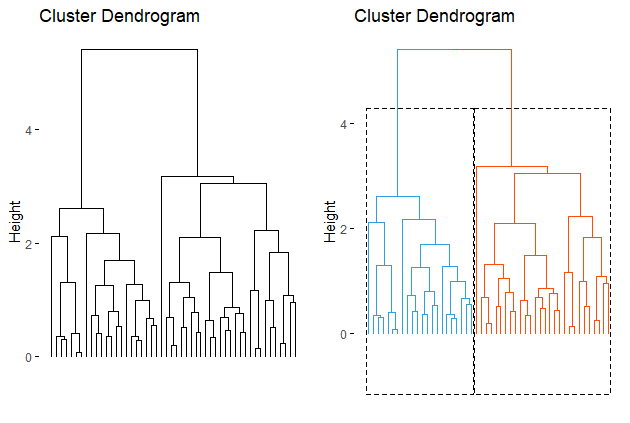}
 \caption{(Left) Dendrogram obtained from 
 hierarchical clustering with average linkage. 
 (Right) By cutting the dendrogram at height 4, 
 two clusters are obtained.}
 \label{fig:dend_ex}
\end{figure}

Hierarchical clustering has some advantages when
compared to partition-based clustering.
For instance, it does not require
the specification of the number of clusters.
Also, as shown in \cref{fig:dend_ex},
a dendrogram displays 
a general similarity structure of the data by
providing a multiresolution view.
The lower in the dendrogram that 
two observations are merged,
the more similar they are.
By using this similarity structure,
\cref{fig:dend_ex} illustrates how
to obtain a partition by
cutting a dendrogram at a given height.

\add{However, there exist 
challenging questions when 
implementing hierarchical clustering. 
For instance, how does one choose 
a particular hierarchical clustering method
among the many that are available? (question 1)
Specifically, how can the performance of 
a dendrogram be evaluated? Also, 
after a clustering method is chosen, 
how do the available features relate to 
the dendrogram? For example,
how can the position of 
each sample in the dendrogram be explained
in terms of its feature values? (question 2)
Also, which features are best segmented by
the dendrogram? (question 3)
}

\textbf{Novelty.} \add{
To the best of our knowledge, our framework is the first one to address the above questions by utilizing the full hierarchical structure provided by the dendrogram. Unlike current approaches, which require the user to generate a partition by cutting the dendrogram at a specified level \citep{Datta2003, Hubert1985, Andrew2007, Yeung2001, kassambara2017, jinwook2002, galili2015, ismaili2014, badih2019}, our approach makes use of the richer structure provided by the dendrogram and does not restrict itself to a single partition, thus avoiding loss of information, as demonstrated in \cref{sec:results}.
 }

\subsection{\add{Overview of the method}}

We answer the questions in the last section 
through \add{probabilistic evolutionary models}
commonly used in phylogenetic analysis \citep{joy2016ancestral,pupko2020gentle,borges2021}.  
The key insight in this framework is 
to view a dendrogram as a phylogeny.
This metaphor is helpful since
the phylogeny \add{is associated to} 
a probabilistic evolutionary model.

\add{Probabilistic evolutionary models describe how each characteristic evolves over a phylogeny. In the context of hierarchical unsupervised learning, this translates into modeling how a feature evolves over a dendrogram using phylogeny information and feature data. These models predict feature values for 
leaves of the dendrogram and also for 
internal nodes, a task known as
ancestral state reconstruction. 
This reconstruction allow one to visualize 
each feature's segmentation by the dendrogram,
thus providing an answer to (2).
Also the evolutionary model's predictions
provide a measure for 
the accuracy of each dendrogram.
By adapting ideas from cross-validation,
we obtain a loss used for dendrogram selection, 
thus answering question (1). Finally, 
feature importance is associated to how well
the probabilistic evolutionary model
predicts that feature, 
which answers question (3).}

\begin{figure}
    \centering
\begin{subfigure}{1\linewidth}
    \centering
 \centering
 \scalebox{0.6}{
 \begin{tabular}{c|ccccc}
  \textbf{Method} & McQuitty & Average linkage  
  & Diana & Ward D &
  Single linkage \\
  \hline
  \textbf{Loss} & \textbf{0.545} & 0.546 & 0.561 & 0.578 & 0.651
 \end{tabular}
 }
 \caption{\add{Cross-validated loss (CVL)} \add{indicates which} hierarchical clustering method \add{to use. In this case, McQuitty has the best performance.}}
 \label{tab:ceramic_samples_scores}
\end{subfigure}
\vspace{2mm}

\begin{subfigure}{1\textwidth}
 \centering
 \includegraphics[scale = 0.45]{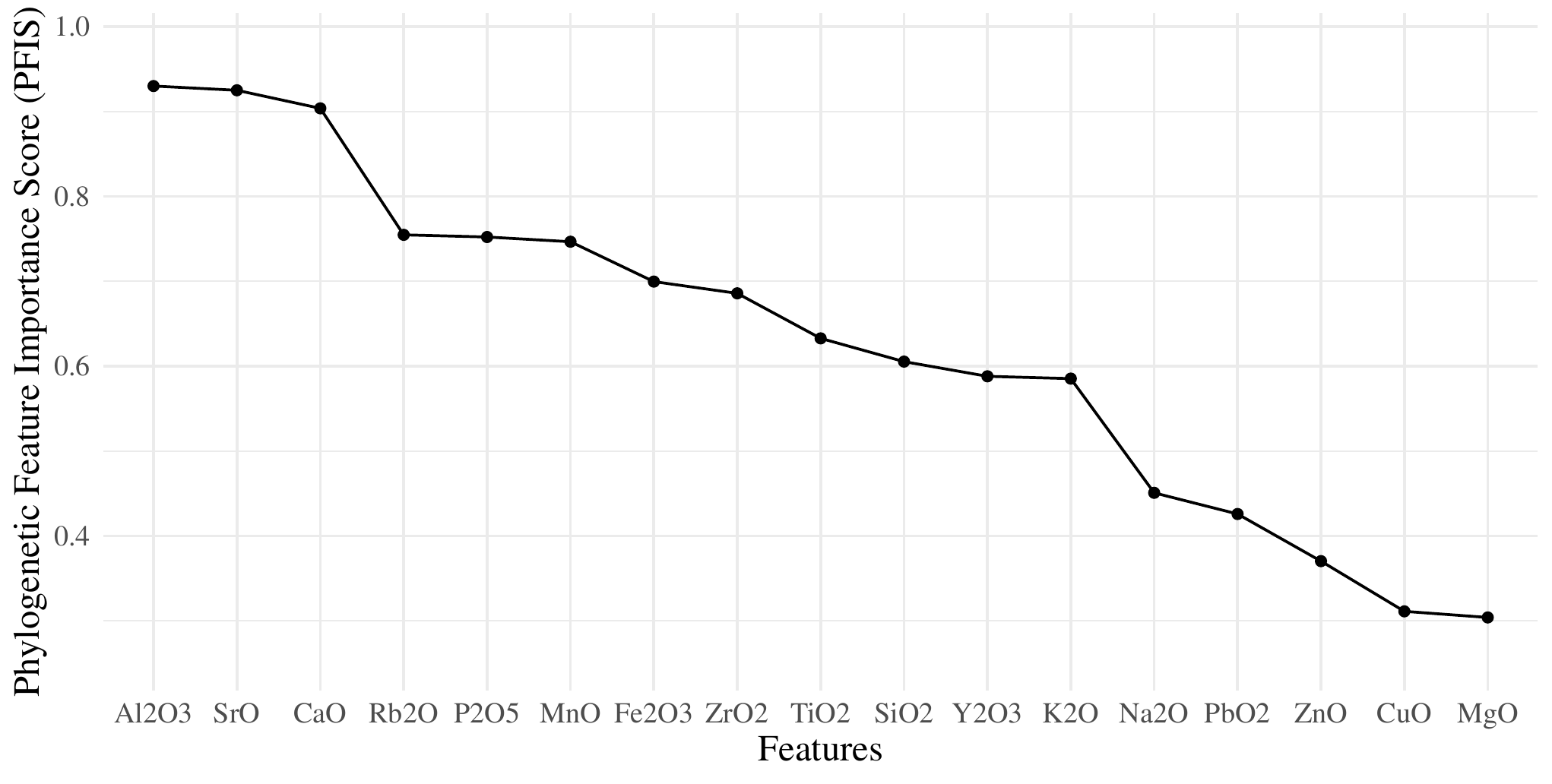}
 \caption{\add{Phylogenetic feature importance score (PFIS) shows the} importance of each feature in Mcquitty dendrogram.}
 \label{fig:toy_ceramic_importance_score}
\end{subfigure}
\vspace{2mm}
\begin{subfigure}{1\textwidth}
 \centering
 \includegraphics[scale = 0.45]{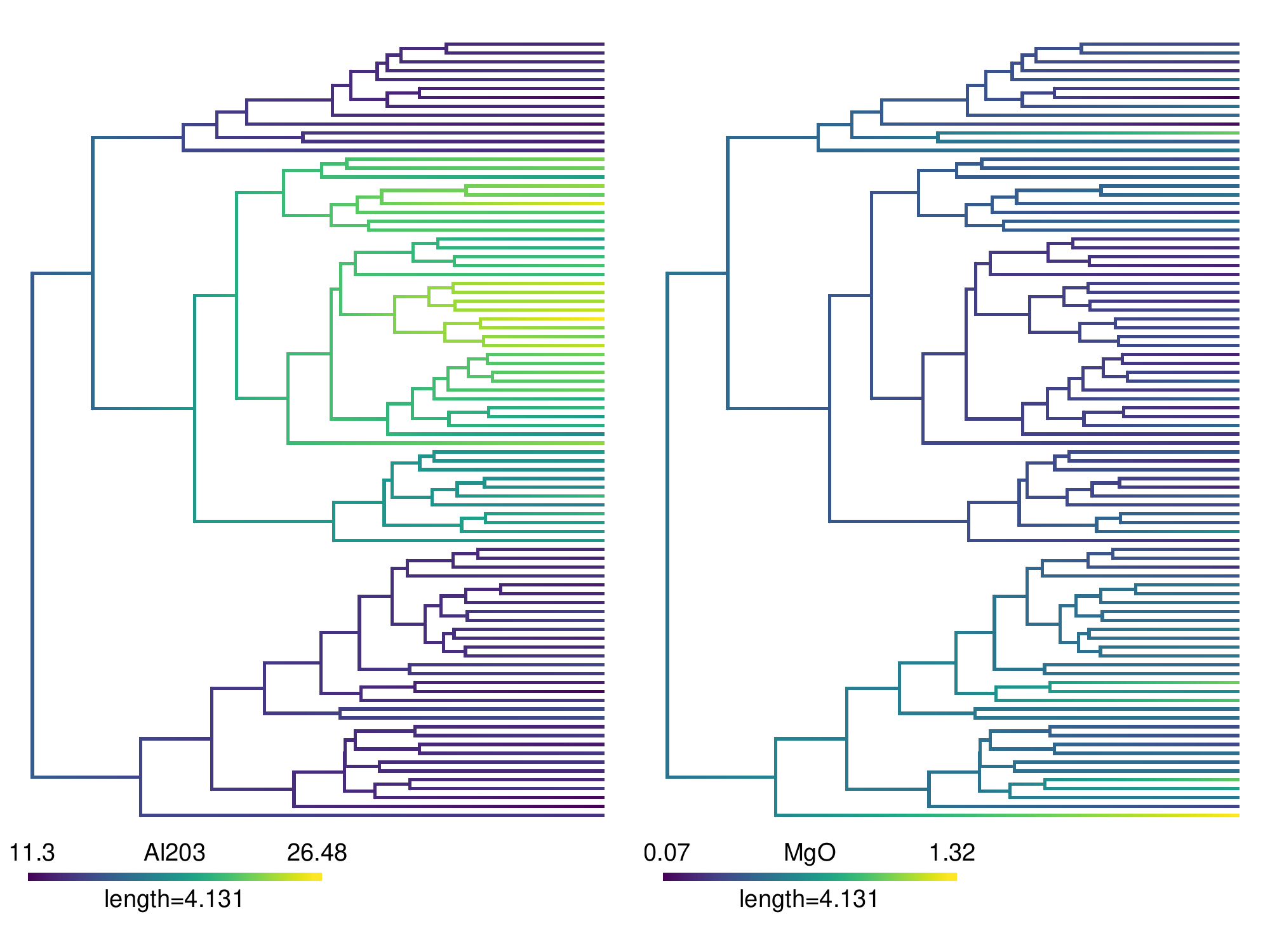}
 \caption{Evolutionary dendrograms \add{gives insights about the clustering by describing
how each feature behaves in each internal node.}}
 \label{fig:toy_ceramic_dendrograms}
\end{subfigure}
\caption{Illustration of the proposed 
hierarchical clustering methodology to 
the ceramic dataset. In panel (a), 
the prediction loss measures how well 
each  hierarchical clustering method explains the data. 
In this example, McQuitty is the best method. 
In panel (b), the feature importance score determines
which of the features are best explained by
the McQuitty dendrogram.
While Al$_2$O$_3$ is the best explained by the dendrogram,  
MgO is the least explained. 
In panel (c), the evolutionary dendrograms 
for Al$_2$O$_3$ (left) and MgO (right) provide insights on \emph{how} the hierarchical structure explains 
the behaviour of each of these feature.}
\end{figure}

We illustate \add{our} approach using 
the \textbf{ceramic samples}
dataset\footnote{\url{https://archive.ics.uci.edu/ml/datasets/Chemical+Composition+of+Ceramic+Samples}}, 
which contains information about 
the chemical composition of \add{88} ceramic samples. \add{This dataset contains 17 features, each related to the percentage of certain chemical compound present in each sample (e.g., AL$_2$O$_3$ (Aluminium Oxide), CaO (Calcium Oxide), etc).}
The table in \cref{tab:ceramic_samples_scores} provides
the hierarhical clustering prediction loss for
a few clustering methods.
This loss provides an immediate way of
comparing the performance of the methods.
For instance, since McQuitty has the lowest loss,
it has the highest performance in this dataset,
closely followed by average linkage. 
Also, \cref{fig:toy_ceramic_importance_score} 
illustrates the feature importance score.
While AL$_2$O$_3$, SrO and CaO are 
the most relevant features in 
the construction of the dendrogram, 
MgO, CuO and ZnO are the least important ones.
Finally, \cref{fig:toy_ceramic_dendrograms} 
illustrates \emph{how} the 
hierarquical structure explains 
the distribution behaviour of each feature. 
The left dendrogram shows that Al$_2$O$_3$ is
well segmented by the dendrogram, as explained by
the even color distribution among its major branches.
That is, while 2 major clusters are characterized by
low values for AL$_2$O$_3$, a third major cluster is
characterized by high values of AL$_2$O$_3$.
The right dendrogram shows that MgO is 
not so well explained by the dendrogram,
since its major branches contain
a high variability of colors.
None of the illustrations required us
to fix the number of clusters.

This work is organized as follows:
\Cref{sec:review} reviews 
hieararchical clustering and \add{probabilistic evolutionary models},
\cref{sec:methods} formulates and 
explains our methodology,
\cref{sec:related_work} outlines related approaches, 
and \cref{sec:results} shows experimental results.

\section{The relation between 
hierarchical clustering and 
probabilistic evolutionary models} 
\label{sec:review}

Hierarchical clustering and phylogenetics share
a common trait: dendrograms and phylogenies are
both formalized as trees. That is,
the difference between a dendrogram and a phylogeny is
given solely by the context in which it is used.
While a dendrogram is used to specify 
the similarity structure between instances,
the phylogeny describes evolutionary relationships.
Since both types of analysis share a common formalism,
it is possible to adapt the methods of one to the other.

Based on this shared formalism,
we also use a common notation for describing
the elements of a tree.
The left panel in \cref{fig:dend_ex}
depicts a tree.
The nodes at the bottom of the tree are called leaves.
These leaves often represent instances in the dataset.
As one moves up the tree, 
the leaves merge into internal nodes, and then
internal nodes merge with one another 
until the root is reached at the top of the tree.
The descendants of a node are 
the leaves that stem from it.

The following subsections briefly review
hierarchical clustering and 
phylogenetical \add{evolutionary models.}

\subsection{Hierarchical methods}

Hierarchical clustering methods yield 
a tree-based representation of the data named dendrogram,
as illustrated in Figure \ref{fig:dend_ex}.
In clustering, a node is often interpreted as the 
group of instances which are its descendants.
The earlier a merge between two nodes, 
the more similar are
the corresponding groups of descendants
\citep{James2013introduction}. 

Using the dendrogram, it is possible 
to partition the instances according to their similarity.
Such a partition is obtained by
cutting the dendrogram with a horizontal line
at a given height, as shown on
the right panel of Figure \ref{fig:dend_ex}.
While cutting at lower heights obtains groups
with a greater similarity,
cutting at higher heights obtains 
a smaller number of groups.
A trade-off between these properties
yields the most interpretable partition.

Hierarchical clustering methods can be 
divided into two  paradigms: 
agglomerative (bottom-up) and 
divisive (top-down) \citep{Elements2009}. 
Agglomerative strategies start at 
the leaves of the dendrogram, iteratively
merging selected pairs of branches until
the root of the tree is reached. 
The pair of branches chosen for merging is 
the one that has the smallest measurement of 
intergroup dissimilarity. 
Divisive methods start at the root at
the root of the tree. Such methods iteratively
divide a chosen branch into smaller ones
until the leaves are obtained.
The splitting criterion maximizes
a measure of between-group dissimilarity.

\subsection{Evolutionary models and ancestral state reconstruction in phylogenetics}

A phylogeny represents 
the evolutionary relationships among instances based on
similarities in their features (traits) \citep{felsenstein1985phylogenies}. 
Each leaf of a tree often represents, for example,
a specimen, a species or a family. 
In this context, an internal node is interpreted as 
the most recent common ancestor of its descendants.

It is often reasonable to posit that
these ancestors have unobserved feature values.
In this context, \emph{ancestral state reconstruction} is
the task of estimating unobserved feature values \add{using \emph{phylogenetic evolutionary models} based on} the \add{phylogeny} and  feature values in the dataset.
This reconstruction is performed by \add{probabilistic evolutionary methods like}
SIMMAP \citep{bollback2006simmap}, for categorical features and by
methods based on Brownian motion
\citep{OMeara2006,Clavel2015,revell2012phytools}, for continuous features. \add{These models can also provide feature value predictions at the leaves.}

\begin{figure}[h]
 \centering
 \begin{subfigure}{.48\textwidth}
  \centering
  \includegraphics[scale = 0.35]{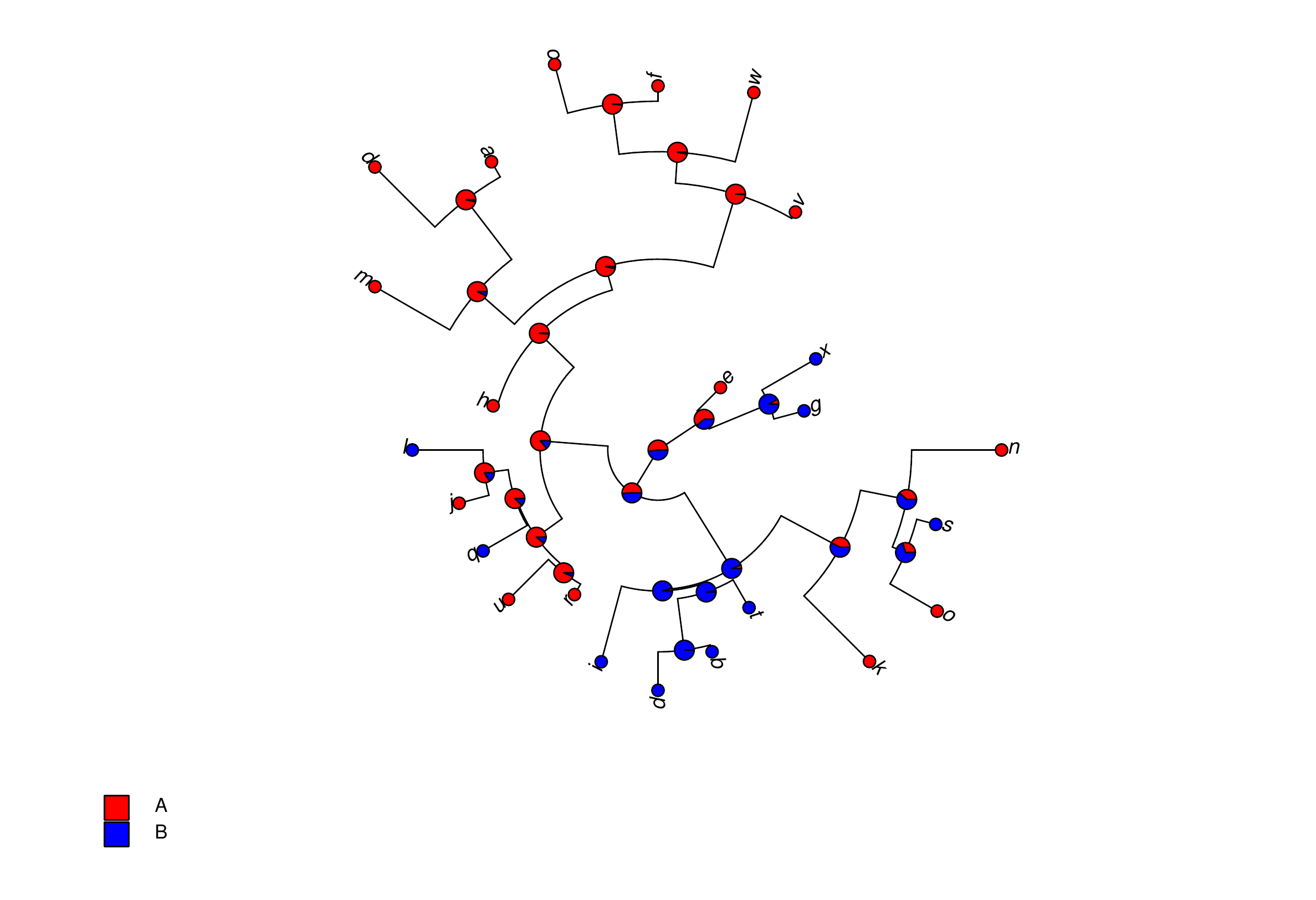}
  \captionof{figure}{Categorical feature}
  \label{fig:phytools_examples_1}
 \end{subfigure}\hspace{2mm}%
 \begin{subfigure}{.48\textwidth}
  \centering
  \includegraphics[scale = 0.475]{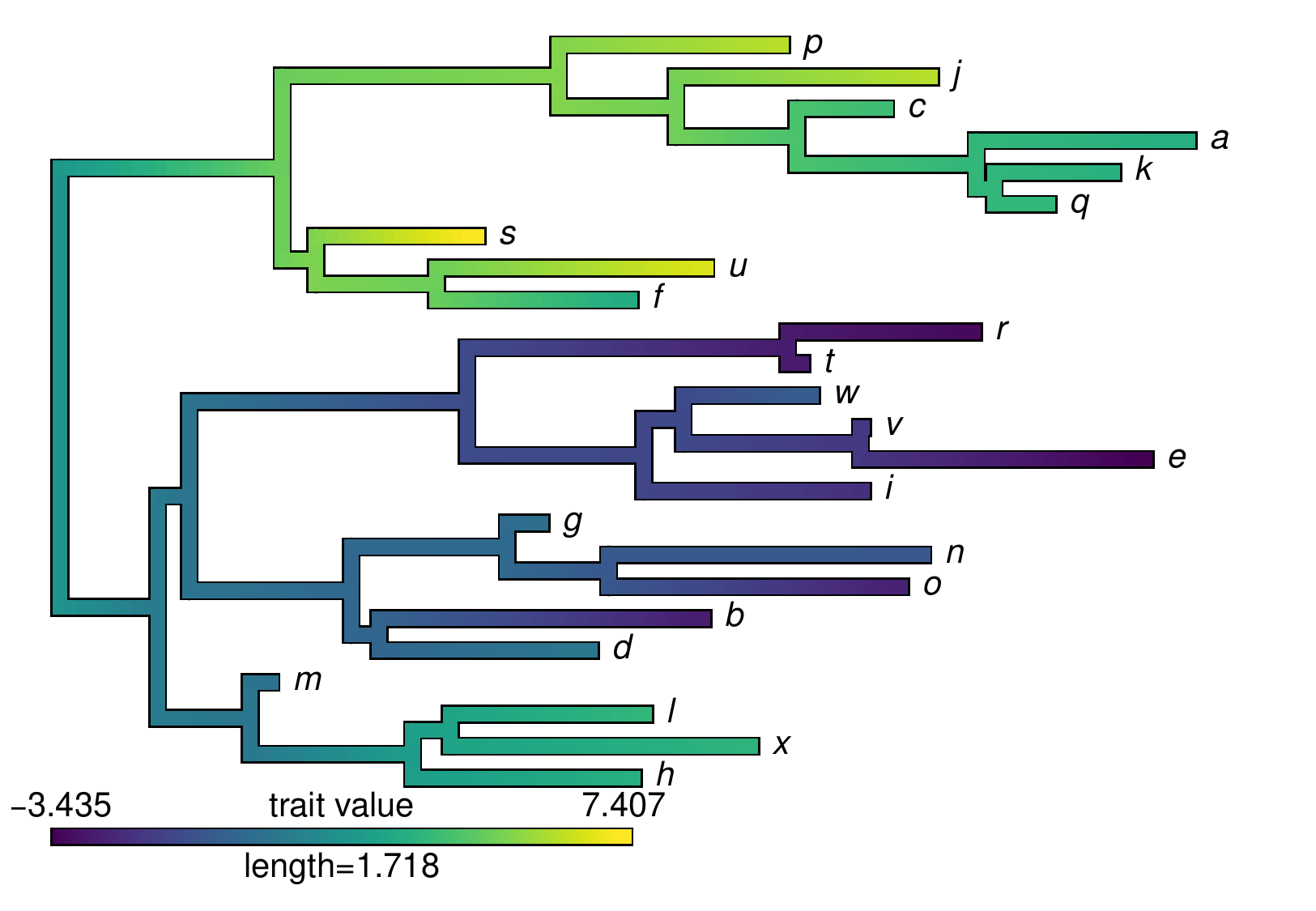}
  \captionof{figure}{Continuous feature}
  \label{fig:phytools_examples_2}
 \end{subfigure}
 \caption{Illustration of ancestral state reconstruction methods.}
 \label{fig:phytools_examples}
\end{figure}

The  \add{ancestral state reconstruction performed by these evolutionary} methods
are illustrated in Figure \ref{fig:phytools_examples}, which
was obtained using the \textit{phytools} package \citep{revell2012phytools}\footnote{\url{http://www.phytools.org/Cordoba2017/ex/15/Plotting-methods.html}}.
Figure \ref{fig:phytools_examples_1} illustrates
ancestral state reconstruction for
a categorical binary feature.
The pie chart over each node is
the estimated probability distribution of 
its feature value.
Figure \ref{fig:phytools_examples_2} illustrates
ancestral state reconstruction for
a continuous feature.
The color of each node represents
the expected value of its feature value.  \add{In this case, the expected value of the feature along the edges can be calculated using a Brownian motion model.}
 
\section{Methodology}
\label{sec:methods}
 
Since both dendrograms and phylogenies are 
formally represented by a tree, 
it is possible to use 
common interpretations and intuition 
of one in the other.
In particular, it is useful to interpret
an internal node of a dendrogram as
an ancestor or an archetype of its descendants.
Using this point of view, 
one can \add{fit evolutionary models to}
hierarchical clustering \add{and perform ancestral state recognition together with leaf prediction}.
The following subsections describe 
some of the benefits of this point of view.
Subsection \ref{subsec:graphical_analysis} shows
useful graphical representations based on
ancestral state reconstruction \add{named \textbf{evolutionary dendrograms}.}
Subsections \ref{subsec:score} and
\ref{sec:importance} show how
this connection leads, respectively,  to
a loss for clustering methods \add{(CVL)}
and to a measure for 
feature importance \add{(PFIS)} in a dendrogram. 

\begin{figure}[h]
    \centering
    \includegraphics[scale = 0.4]{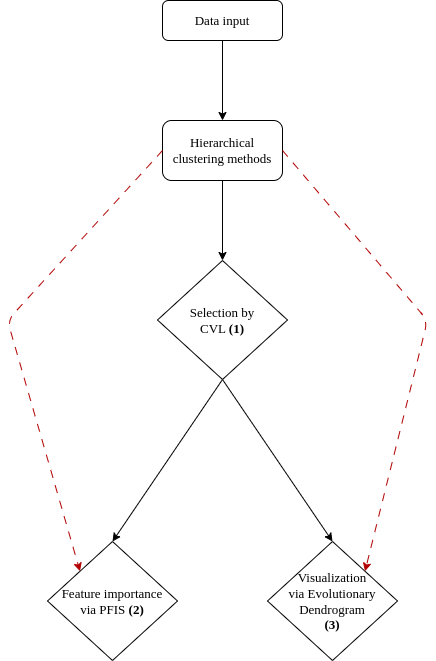}
    \caption{\add{Flowchart of our full (only continuous arrows) and partial (dashed arrows) framework. The numbers displayed in parenthesis show which question each step answers. The full framework selects the best performing hierarchical clustering  method according to our proposed loss (CVL). Afterwards, both feature importance (PFIS) and evolutionary dendrograms are plotted to provide insights and to better explain the dendrogram that was selected by CVL. If we already have a single hierarchical method, the selection step  can be skipped and we can  proceed directly to plotting feature importance and evolutionary dendrograms, as indicated by the dashed arrows.}} \label{fig:flowchart_hierarchical_clustering}
\end{figure}

\add{These tools can be combined into a complete hierarchical clustering analysis framework as shown in the flowchart of Figure \ref{fig:flowchart_hierarchical_clustering}.  First, we fit all hierarchical clustering methods we desire based on the input data. 
Next, for each dendrogram, 
an evolutionary model predicts
each feature's values through a
leave-one-feature-out strategy.
These predictions yield a loss (CVL)
for each dendrogram, thus allowing
the selection of the one with best perfomance.
For the selected dendrogram, 
we compute feature importance score 
using PFIS. This score uses 
the evolutionary model and 
a leave-one-observation-out strategy to
determine how well each feature is
explained by the dendrogram.
We also visualize how the chosen dendrogram
explains each feature using 
the probabilistic evolutionary model.
This model uses 
ancestral state reconstruction to plot 
how each feature evolves 
over the dendrogram.}

\add{In what follows, we denote the dataset by $$\mathbb{D}= \left[ x_{i,j} \right]_{i=1,\ldots,n}^{j=1,\ldots,p},$$
where $i$ indexes the instances, and $j$ indexes the features. Also, we denote a dendrogram by $\T$. When we want to emphasize that $\T$ is built using data $\mathbb{D}$, we 
denote it by $\hat{\T}$.}

\subsection{Evolutionary dendrograms} \label{subsec:graphical_analysis}

By interpreting a dendrogram as a phylogeny,
one can \add{fit a probabilistic evolutionary model to the dendrogram and} obtain insights about the clustering through
ancestral state reconstruction.
This approach offers 
a multiresolution visualization of the data,
which does not require a pre-specified number of clusters.
More specifically, this approach describes how 
each feature behaves in each internal node of the dendrogram.
Such a visualization provides insight about
what partition size yields 
a more useful description of the data.  

\add{Because ancestral state reconstruction visualization tools are already implemented in phylogenetic analysis packages like \textit{phytools} \citep{revell2012phytools} and \textit{ape} \citep{paradis2019ape}  both for continuous and categorical traits, it is straightforward to apply these methods to the hierarchical clustering context: one only needs to take the resulting dendrogram as the input for these procedures. We provide code for this simple adaptation in \url{https://github.com/Monoxido45/PhyloHclust} and illustrate this graphical analysis with
an application to 
the USArrests \citep{USArrests}
and Iris \citep{Fisher} datasets. }

\subsubsection{Evolutionary dendrograms applied to 
the USArrests dataset}
\label{sec:usarrests}

\begin{figure}[!hppt]
 \centering
 \begin{subfigure}{1\textwidth}
  \centering
 \includegraphics[scale=0.45]
 {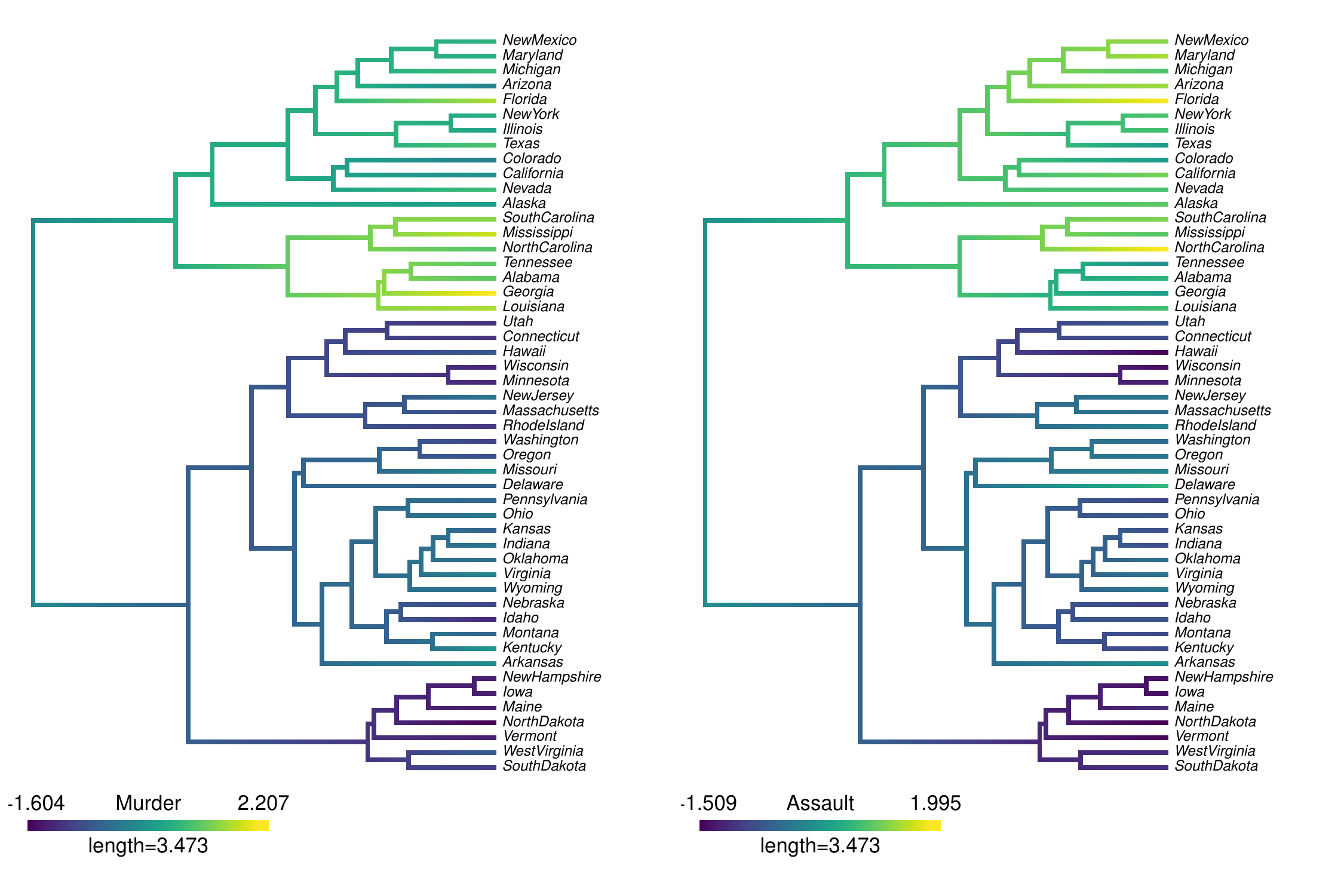}
  \label{fig:mcquitty_murder_assault_dend}
 \end{subfigure} \vspace{2mm}%
 \begin{subfigure}{1\textwidth}
  \centering
  \includegraphics[scale = 0.45]
 {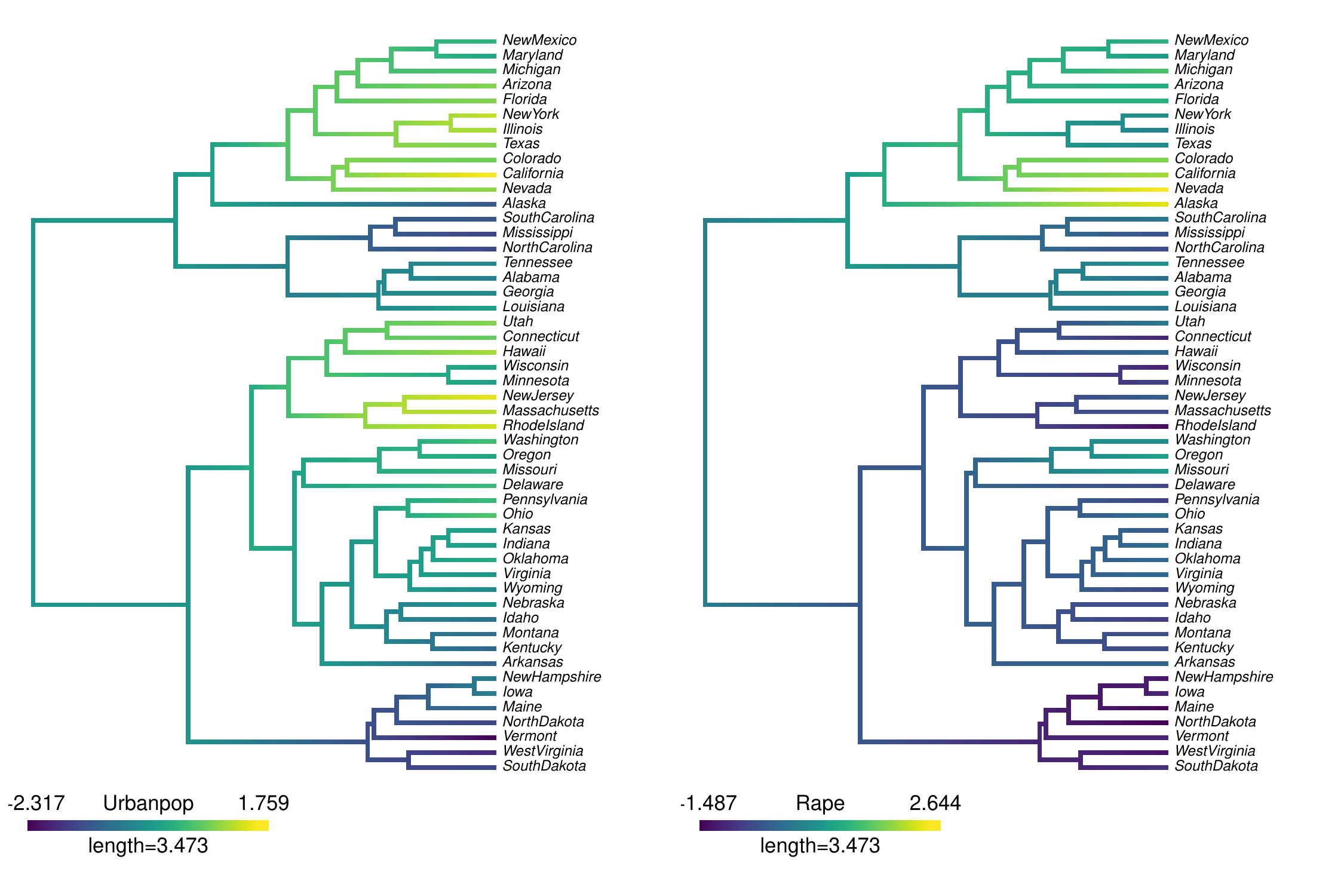}
  \label{fig:mcquitty_urbanpop_rape_viridis}
 \end{subfigure}
 \caption{Evolutionary dendrograms for 
 standardized \textbf{Murder} (top left), \textbf{Assault} (top right) and \textbf{Rape} (bottom right) rates per 100.000 residents and  standardized \textbf{Urban population} (bottom left) applied to a clustering obtained using the McQuitty method.}
 \label{fig:mcquitty_assault_murder_urbanpop_rape_viridis}
\end{figure}

The USArrests dataset \add{\citep{USArrests}} describes the urban population and
murder, assault and rape rates per 100,000 residents for
each of the USA states.
Using this data, we obtain 
a dendrogram for the USA states using
the McQuitty method and 
perform ancestral state reconstruction applying a \add{fitted Brownian Motion evolution model} along the edges
\citep{felsenstein1985phylogenies, revell2012phytools, revell2013two}. \add{Figure \ref{fig:mcquitty_assault_murder_urbanpop_rape_viridis} shows
the evolutionary dendrograms for standardized \textbf{Murder} (top left), \textbf{Assault} (top right) and \textbf{Rape} (bottom right) rates per 100.000 residents and standardized \textbf{Urban population} (bottom left). Through the evolutionary dendrogram, we can obtain valuable insights, such as the following:}
\add{\begin{itemize}
    \item Since the major branches of 
the top dendrograms have similarly colored descendants in both cases, both \textbf{Assault} and \textbf{Murder} features are well segmented by the dendrogram.
 \item In the top left dendrogram, 
the major internal nodes vary from   high (yellow) to low murder rates (dark blue),
with high \textbf{Murder} rates descendant states listed from South Carolina to Louisiana,
medium murder rates (green) from New Mexico to Alaska, mild to low rates (light blue) from Utah to Arkansas, low rates (dark blue) from New Hamsphire to South Dakota.
\item In the right dendrogram, 
\textbf{Assault} rates are segmented
in a similar way as \textbf{Murder}'s dendrogram. The main difference from the previous analysis is that
high assault rates are not well separated from
medium assault rates.% Indeed,
%the internal nodes in the list of states
%from New Mexico to Louisiana remain
%light green until close to the leaves.
\item One might use a coarse partition which
divides the states into three groups:
one with medium to high
murder and attack rates,
another with mild rates, and 
the last with low rates. Furthermore, one could also refine 
the partition by dividing
the first group into
two others with respectively 
high and medium murder rates.
\item The deep internal nodes of \textbf{Rape} rate dendrogram are
colored similarly to \textbf{Assault} rate dendrogram.
That is, these features are segmented similarly 
by the dendrogram. 
\item  \textbf{Urban population} is not
well segmented by the dendrogram: the major branches of the \textbf{Urban Population} dendrogram have differently colored descendants, and almost all of its deep internal nodes are green. Thus, we could obtain more internally homogeneous groups by selecting internal nodes that are closer to the leaves, but these groups would be similar to others that are not in the same major branch. For instance, we can select internal nodes in the first major branch with descendants listed from South Carolina to Lousiana and internal nodes in the second major branch with descendants listed from New Hamsphire to South Dakota.
\end{itemize}
}

\add{It would be much harder to perform the above analysis directly through a 
standard partition-based boxplot analysis. 
One would need to cut the dendrogram
at several different heights and then 
gather the collection of boxplots obtained
for each feature and heigth. 
By doing so, the multi-resolution view 
offered by the dendrogram is lost and 
clustering cannot be visualized 
at an individual level. 
We further detail 
this comparison in 
Section \ref{sec:ex_evo_dendro}}.

\begin{comment}
\begin{figure}[hp!]
 \centering
 \includegraphics[scale=0.44]
 {figures/mcquitty_assault_and_murder_viridis.pdf}
 \caption{Evolutionary dendrograms for 
 standardized \textbf{Murder} (left) and 
 \textbf{Assault} (right) rates per 100.000 residents
 applied to a clustering obtained using the McQuitty method.}
 \label{fig:mcquitty_murder_assault_dend}
\end{figure}

\begin{figure}[hp!]
 \centering
 \includegraphics[scale = 0.44]
 {figures/mcquitty_urbanpop_rape_viridis.pdf}
 \caption{Evolutionary dendrograms for 
 standardized \textbf{Urban population} (left) 
 and \textbf{Rape} rate (right) 
 applied to a clustering obtained using the McQuitty method.}
 \label{fig:mcquitty_urbanpop_rape_dendrogram}
\end{figure}

\end{comment}

\subsubsection{Evolutionary dendrograms applied to 
the Iris dataset}
\label{sec:iris}

The Iris dataset \add{\citep{Fisher}} contains measurements related to $150$ flowers.
Specifically, it contains the width and length of
the sepals and petals of flowers which are classified
into three species: \textbf{setosa}, \textbf{versicolor},
and \textbf{virginica}. In this analysis, we performed
clustering using complete linkage 
over the petal and sepal measurements.

\begin{figure}[!hppt]
 \centering
 \includegraphics[width= 0.7\textwidth]
 {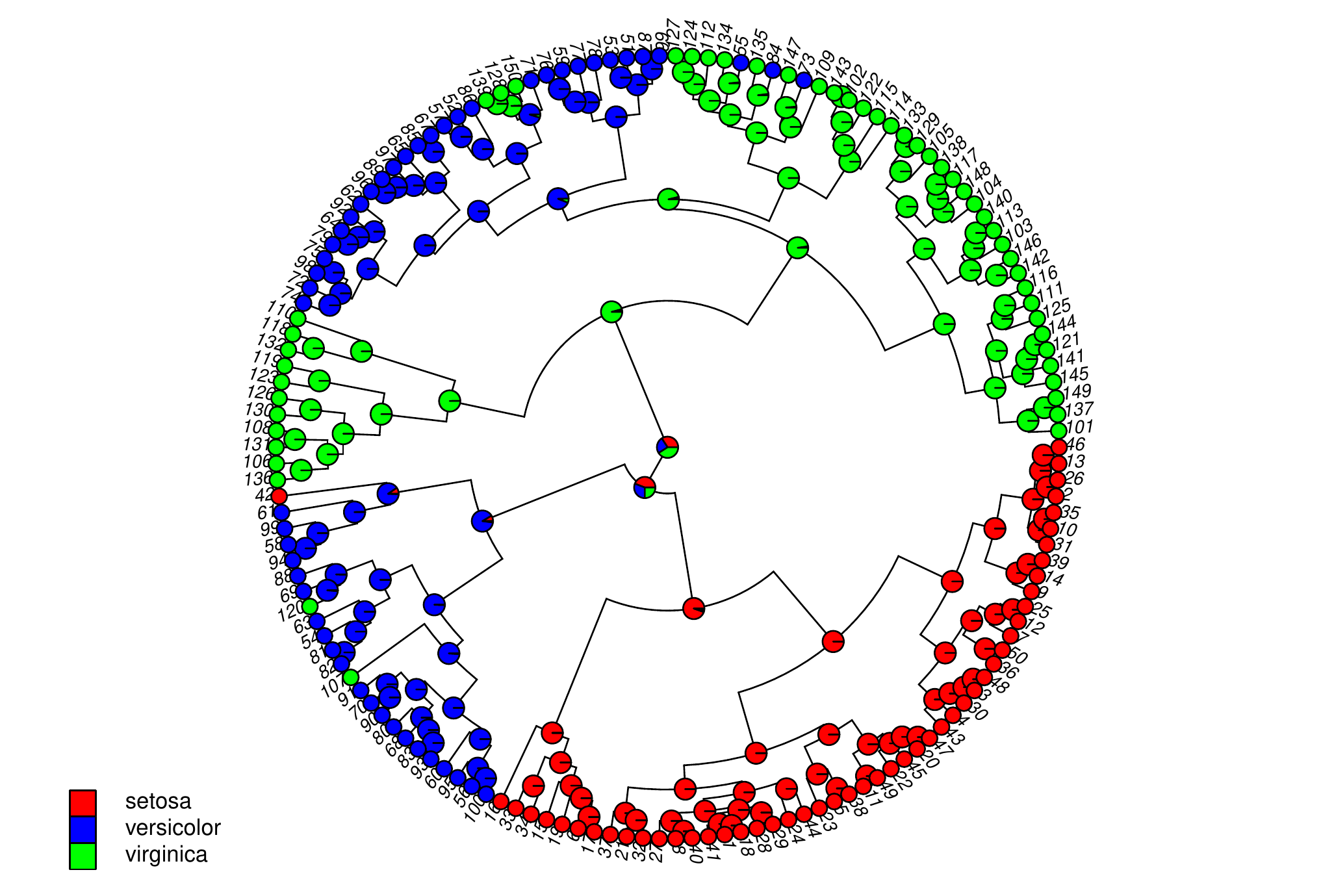}
 \caption{Evolutionary dendrogram obtained by applying acenstral state reconstruction of species based on the SIMMAP algorithm. The original dendrogram is obtained by performing complete linkage clustering to petal and sepal widths and heights.}
 \label{fig:iris_example}
\end{figure}

Although the species were not used in the clustering,
a graphical analysis shows how well the
resulting dendrogram segments this label.
Figure \ref{fig:iris_example} illustrates
the evolutionary dendrogram for species using
SIMMAP \citep{bollback2006simmap}.
The Figure shows that,
using a partition of size $5$,
the dendrogram segments the label adequately.

\subsection{Cross-validated loss (CVL)} 
\label{subsec:score}

Before one can interpret a dendrogram,
it is necessary to choose a clustering algorithm.
In order to do this, we propose a new loss function for
evaluating the performance of each algorithm.

The connection between dendrograms and
phylogenies can also be explored for
choosing between clustering algorithms.
The main advantage of this relation is that
a phylogeny induces predictions over instances.
As a result, it is possible to
adapt to clustering the methods for 
model selection that are used in
supervised learning.
More specifically, our proposal is
an adaptation of a leave-one-feature-out cross-validation
approach for selecting phylogenies \citep{borges2021}. \add{Let $\hat{\T}$ be a dendrogram fitted according to an arbitrary method. The score is given by
$$\mbox{CVL}(\hat{\T}) 
  := p^{-1}\sum_{j=1}^p L_j(\hat{\T}_{-j}),$$
  where
$$L_j(\hat{\T}_{-j}) := 
   n^{-1}\sum_{i=1}^n d_j(\hat{x}_{i,j}, x_{i,j}),$$
   $\hat{\T}_{-j}$ is the dendrogram fitted according to the same method, but without the $j$-th feature,  $\hat{x}_{i,j}$ is 
   a prediction $\hat{\T}_{-j}$ gives to 
$x_{i,j}$, the $j$-th feature of 
the $i$-th instance, and
$d_j(\hat{x}_{i,j}, x_{i,j})$ as the
inaccuracy of $\hat{x}_{i,j}$ 
with respect to $x_{i,j}$.
}
The proposal is \add{summarized} in Algorithm \ref{alg:cvl}.
For Algorithm \ref{alg:cvl} to be operational,
it is necessary to choose a prediction function and
a measure of inaccuracy of predictions.
The choice of these functions depends on
whether the feature under analysis is 
categorical or continuous.

\begin{algorithm}
 \caption{ \small Cross-validation loss (CVL)}
 \label{alg:cvl}
 \algorithmicrequire \ {\small 
 Clustering algorithm, $\hat{\T}$, and
 data $x_{i,j}$, for $1 \leq i \leq n$ and 
 $1 \leq j \leq p$.} \\
 \algorithmicensure \ {\small 
 CVL($\hat{\T}$), the 
 cross-validated predictive loss of $\hat{\T}$.} \\
 \begin{algorithmic}[1]
  \For{each feature, $j$,}
   \State Let $\hat{\T}_{-j}$ be
   obtained by applying $\hat{\T}$ to 
   the data with feature $j$ removed.
   \For{each instance, $i$,}
    \State Let $\hat{x}_{i,j}$ be a prediction for
    $x_{i,j}$ based on $\hat{\T}_{-j}$ and on
    $x_{i',j}$ for $i' \neq i$.
    \State Let $d_j(\hat{x}_{i,j}, x_{i,j})$ measure
    how inaccurate prediction $\hat{x}_{i,j}$ is
    for $x_{i,j}$.
   \EndFor
   \State Let $L_j(\hat{\T}_{-j}) := 
   n^{-1}\sum_{i=1}^n d_j(\hat{x}_{i,j}, x_{i,j})$
   \Comment{Average inaccuracy for $j$}
  \EndFor
  \State Let CVL$(\hat{\T}) 
  := p^{-1}\sum_{j=1}^p L_j(\hat{\T}_{-j})$
  \Comment{$\hat{\T}$'s average inaccuracy}
 \end{algorithmic}
\end{algorithm}

\begin{figure}
    \centering
    \includegraphics[scale = 0.35]{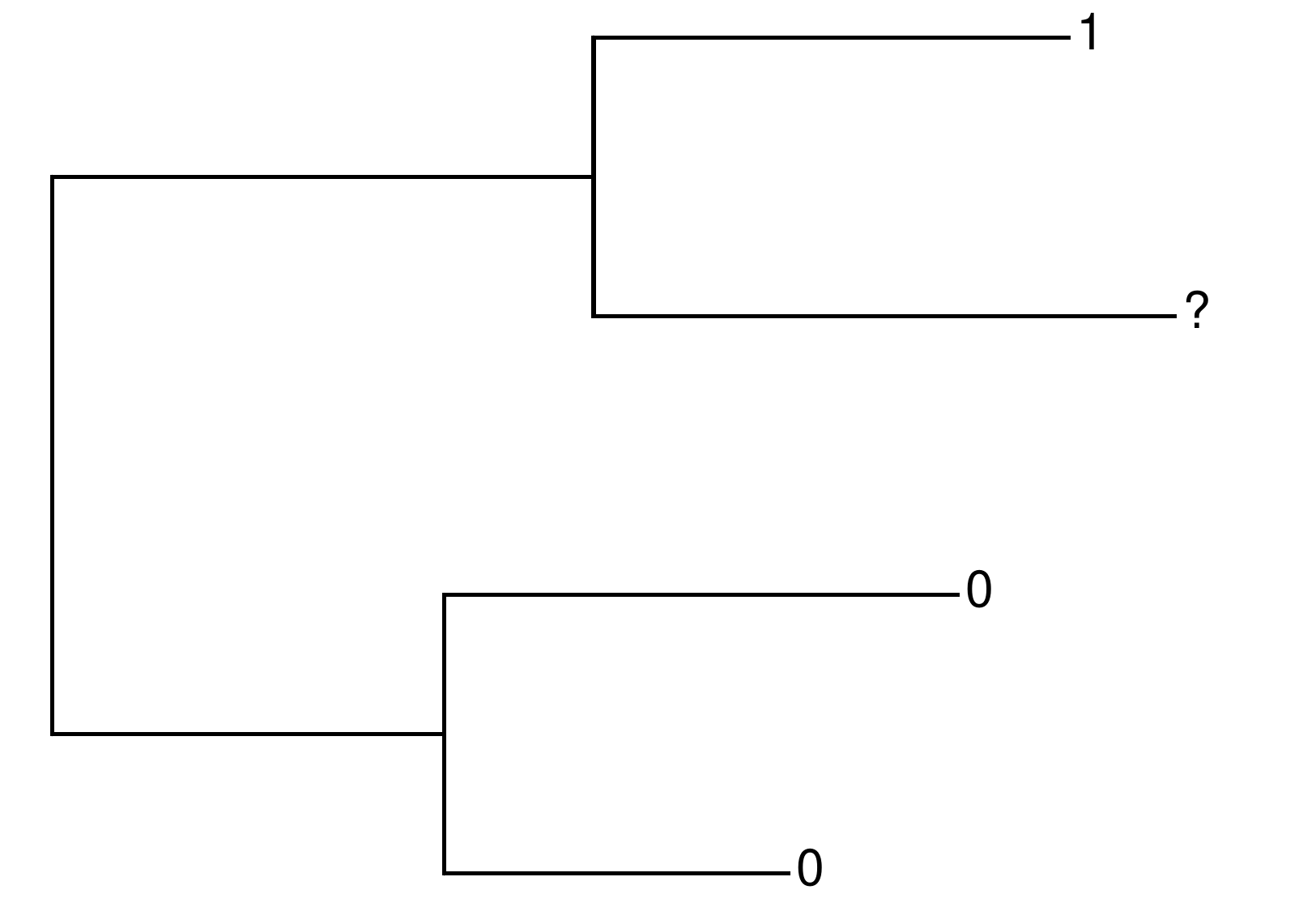}
    \caption{\add{Example of a phylogeny with missing character on a leaf. Evolutionary models  such as SIMMAP predict the value of the feature indicated with a question mark based on the phylogeny (dendrogram) and the other known characters.}}
    \label{fig:example_ancestral_states_recons}
\end{figure}

When the feature is categorical,
we consider that the prediction is
a list of probabilities for each category.
Formally, if $x_{i,j} \in \mathcal{A}$, then
$\hat{x}_{i,j} = (\pi_{i,j,a})_{a \in \mathcal{A}}$,
where $\pi_{i,j,a}$ is 
an estimate of the probability that $x_{i,j} = a$.
Determining these probabilities based on the dendrogram is
analogous to inferring missing characters on
phylogenies, as performed for example by SIMMAP. \add{As illustrated in Figure \ref{fig:example_ancestral_states_recons}, the main goal of such evolutionary models
is to predict  feature values that are missing  using the data from other leaves and the information of the phylogeny (in our case, the dendrogram).}

Hence, we propose using SIMMAP for computing $\pi_{i,j,a}$.
Using such predictions, $d_j$ is defined as
the Brier score, that is,
\begin{align*}
 d_j\left((\pi_{i,j,a})_{a \in \mathcal{A}}, x_{i,j}\right) 
 &= |\mathcal{A}|^{-1}\sum_{a \in \mathcal{A}}
 (\I(x_{i,j} = a) - \pi_{i,j,a})^2.
\end{align*}
When the testing feature is continuous,
we consider that the prediction is
an estimate of a typical feature value.
Drawing from the phylogenetical analogy,
$\hat{x}_{i,j}$ is computed using the 
Brownian Motion process 
\citep{OMeara2006, Clavel2015}, which was
originally developed to infer 
continuous features for 
internal nodes and leaves of a phylogeny. 
In this case, each instance $x_{i, j}$ is standardized (that is, it is normalized to have mean zero and variance one)  before fitting the Brownian Motion process, and $d_j$ is defined as the squared error, 
\begin{align}
    \label{eq:squared_error}
d_j(x_{i,j}, \hat{x}_{i,j}) = (x_{i,j} - \hat{x}_{i,j})^2.
\end{align}

%Besides motivating the above loss for
%clustering algorithms,
%the predictive approach to dendrograms
%can also be used to build 
%a score of feature importance.
%This score is discussed in the next subsection.

\subsection{Phylogenetic feature importance score (PFIS)}
\label{sec:importance}

Based on the idea that dendrograms \add{equipped with an evolutionary model} provide
predictions over their leaves,
it is \add{also} possible to construct
a score for feature importance.
Let $\hat{x}_{i,j}(\hat{\T})$ be a prediction for
$x_{i,j}$ based on \add{the fitted dendrogram $\hat{\T}$} and on
$x_{i',j}$, for $i' \neq i$\add{, obtained using an evolutionary model as in the last section.}
%We define a measure of $\T$'s inaccuracy
%in predicting $j$, $L_j(\T)$, as
%can measure the inaccuracy of $\T$'s 
%in predicting $j$, by
%\begin{align*}
% L_j(\T) &= n^{-1}\sum_{i=1}^n d_j(\hat{x}_{i,j}(\T), x_{i,j}).
%\end{align*}
We 
define the Phylogenetic Feature Importance Score (PFIS) 
of feature $j$ over \add{$\hat{\T}$},
$\mathcal{I}_j(\hat{\T})$, as:
\begin{align*}
  \mathcal{I}_j(\hat{\T})
  &= 1 - \frac{\sum_{i=1}^n d_j(\hat{x}_{i,j}(\hat{\T}), x_{i,j})}{n},
\end{align*}
where $n^{-1}\sum_{i=1}^n d_j(\hat{x}_{i,j}(\hat{\T}), x_{i,j})$ measures the \emph{inaccuracy} of $\hat{\T}$'s 
in predicting $j$.
Notice that, for standardized continuous features and the squared error (Equation \ref{eq:squared_error}), $L_j(\hat{\T})$ is exactly the ratio between sum of squares of residuals and total sum of squares, with $SS_{res} = \sum_{i=1}^n d_j(\hat{x}_{i,j}(\hat{\T}), x_{i,j})$ and $SS_{tot} = n$.
It follows that, in this case, $\mathcal{I}_j(\hat{\T})$ is a coefficient of determination, $R^2$, often used to evaluate the explanation power of statistical models in regression analysis \citep{neter1996applied}.

The main purpose of PFIS is to filter 
features according to how relevant they are in 
the segmentation of the dendrogram. For instance,
the features in Subsection \ref{subsec:graphical_analysis} 
with small importance scores lead to
evolutionary dendrograms without a clear structure.
By knowing this in advance, 
one focus on looking at 
the more informative figures.

\section{Related Work}
\label{sec:related_work}

To the best of our knowledge, 
our  framework to graphical analysis  and  feature importance is 
the first one that uses the 
full hierarchical structure given by a dendrogram.
All previous approaches involve mapping
the dendrogram to a partition of instances\add{, as we detail in the next subsections.}

\subsection{\add{Graphical analysis}}

\add{
Graphical analysis is a commonly used method for evaluating clustering algorithms with a fixed number of groups. This typically involves plotting the distribution of each feature on each cluster using various visualization techniques such as boxplots, barplots, histograms, or radar charts 
\citep{galili2015,kassambara2017,jinwook2002}. }

\add{Another method for visualizing clusters is by creating a scatter plot of the features projected into a low-dimensional space using techniques such as principal component analysis \citep{hair2009multivariate} or t-SNE \citep{van2008visualizing}).  For a more in-depth review of these techniques and their variations, refer to \citet{hennig2015clustering}.}

 \add{It is worth noting that these visualization tools can also be used for hierarchical clustering methods, however, mapping the dendrogram into a partition can lead to the loss of important information(see Section \ref{sec:ex_evo_dendro} for a comparison). }

\add{We are not aware of visualization techniques specifically designed for hierarchical clustering methods that allow for the interpretation of how features vary across multiple resolution clusters.}

\subsection{\add{Feature importance}}

\add{Several feature importance metrics have been developed for specific clustering algorithms. For instance, \citet{FeatureImpCluster} proposes a solution for $k$-means clustering, while \citet{zhu2018variable} presents a framework for determining feature importance in model-based clustering. Similarly, \citet{badih2019} introduce  a method for assessing feature importance in decision tree clustering.}

\add{However, there exist only a few 
feature importance scores that 
can be applied to arbitrary 
partition-based clustering algorithms.
For instance, \citet{ismaili2014} 
trains a classifier that 
predicts the cluster of each instance based
on its feature values.
The feature importance for the clustering is
identified with the feature importance for
the classifier that was trained.
For example, one might use
the mean decrease accuracy for Random Forests \citep{breiman2001random}, 
the absolute coefficient for lasso \citep{tibshirani1996regression}, or other importance metrics used in supervised learning \citep{hastie2009elements,izbicki2017converting,coscrato2019nls,shimizu2022model}.  \citet{badih2019} is closer to 
our approach, and is also based on 
a leave-one-variable out approach (LOVO). However, the feature importance is 
based on computing a 
within-cluster heterogeneity measure.}

\add{These approaches can only be used for
hierarchical clustering by 
mappingthe dendrogram into a partition, 
which leads to loss of information. 
We compare the above approaches 
against our method 
in Section \ref{sec:pfisRF}.}

\subsection{\add{Clustering scores}}

\add{Numerous metrics have been developed 
to assess the performance of 
partition-based clustering. 
Among such metrics are 
the non-overlap score \citep{Datta2003}, 
average distance between means \citep{Datta2003}, 
average distance \citep{Datta2003}, 
adjusted Rand index \citep{Hubert1985},
V-measure \citep{Andrew2007}, 
Silhouette coefficient \citep{Peter1987}, and
Calinski-Harabasz index \citep{Calinski1974}.
 A comprehensive review of these metrics can be found in  \citet{vendramin2010relative}. However, to the best of our knowledge, the only metric that performs cross-validation in a predictive framework, is  Figure of Merit (FOM; \citealt{Yeung2001}), which we review in Section \ref{sec:testsCVL}. 
 
 In order to apply 
 the above approaches to
 a hierarchical clustering, 
 one must map the dendrogram into a partition. Therefore, two dendrograms that 
 are mapped to the same partition
 have the same score. 
 For this reason, as demonstrated 
 in Section \ref{sec:testsCVL}, 
 CVL commonly outperforms FOM.}

\add{A metric that is similar in concept to CVL is the cophenetic correlation coefficient \citep{sokal1962comparison, saraccli2013comparison, timofeeva2019evaluating}. 
This coefficient assesses the accuracy of the entire hierarchical structure by borrowing ideas from phylogenetic analysis. It measures the correlation between the euclidean distance between instances and the cophenetic distance induced by the phylogeny. However, it does not employ cross-validation, which can result in overfitting.}

\add{Section \ref{sec:testsCVL} compares CVL to the cophenetic correlation coefficient, Figure of Merit and Adjusted Rand Index. }

\section{Experiments}
\label{sec:results}

In this section, we compare our proposal for
graphical analysis, feature importance and
clustering loss to other methods based on
partition clustering.
In order to apply the latter methods,
partitions are generated by cutting
the dendrogram at varying heights.

\subsection{Evolutionary dendrograms vs. 
grouped boxplots}
\label{sec:ex_evo_dendro}

Boxplots are a common way of visualizing 
how a feature is segmented by a dendrogram.
First, a partition is obtained by cutting
the dendrogram at a given height.
Next, a boxplot is drawn
for each cluster in the partition.
By successively cutting the dendrogram at
different heights, one can 
fully visualize how the feature is segmented.
However, much of the information in the dendrogram is lost in this procedure.

\begin{figure}[!hppt]
 \centering
 \begin{subfigure}{0.45\textwidth}
  \centering
  \includegraphics[width=\textwidth]
  {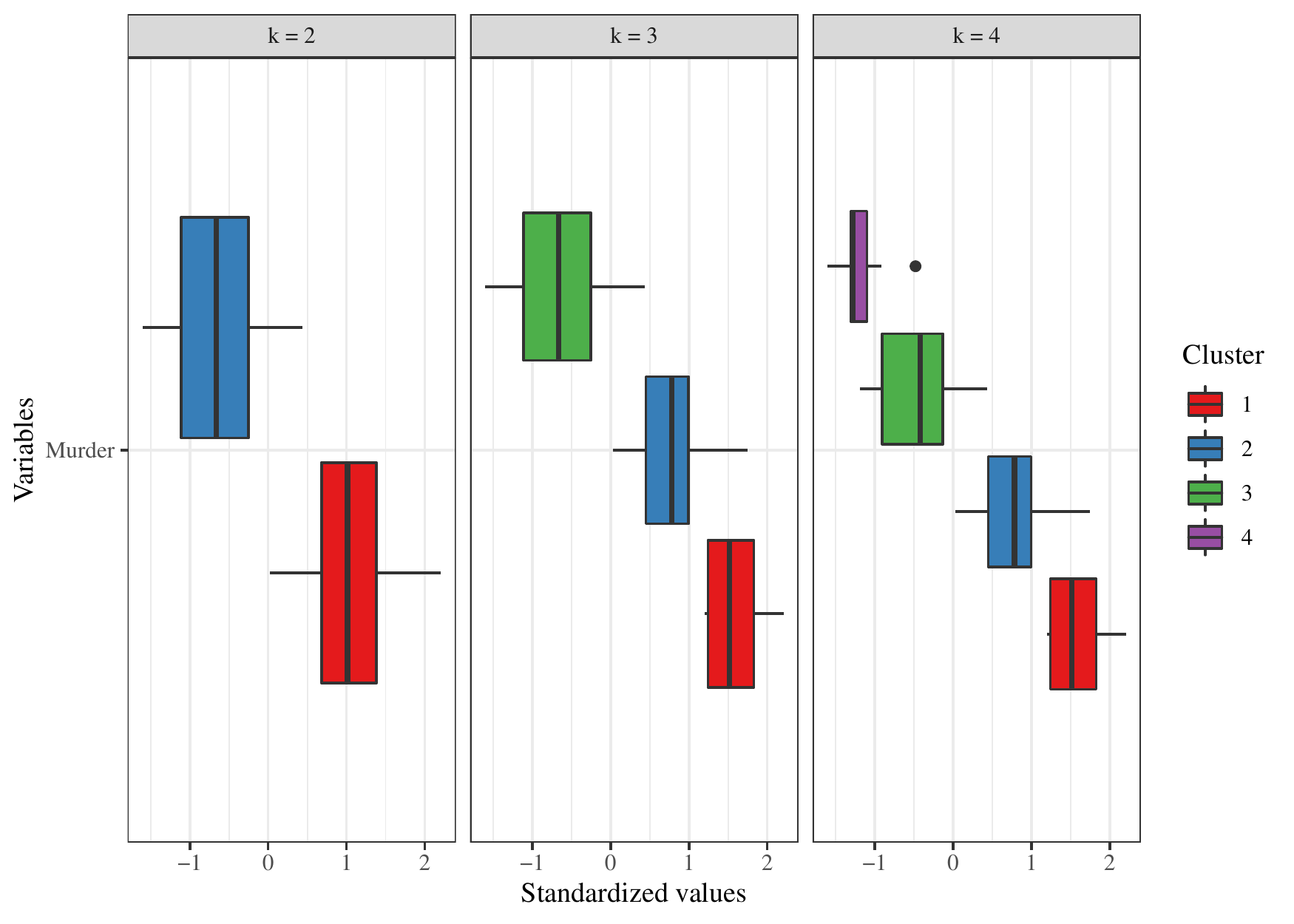}
  \captionof{figure}{Boxplots with different segmentation}
  \label{fig:boxplots_usarrests}
 \end{subfigure}\hspace{2mm}%
 \begin{subfigure}{0.45\textwidth}
  \centering
  \includegraphics[width=\textwidth]
  {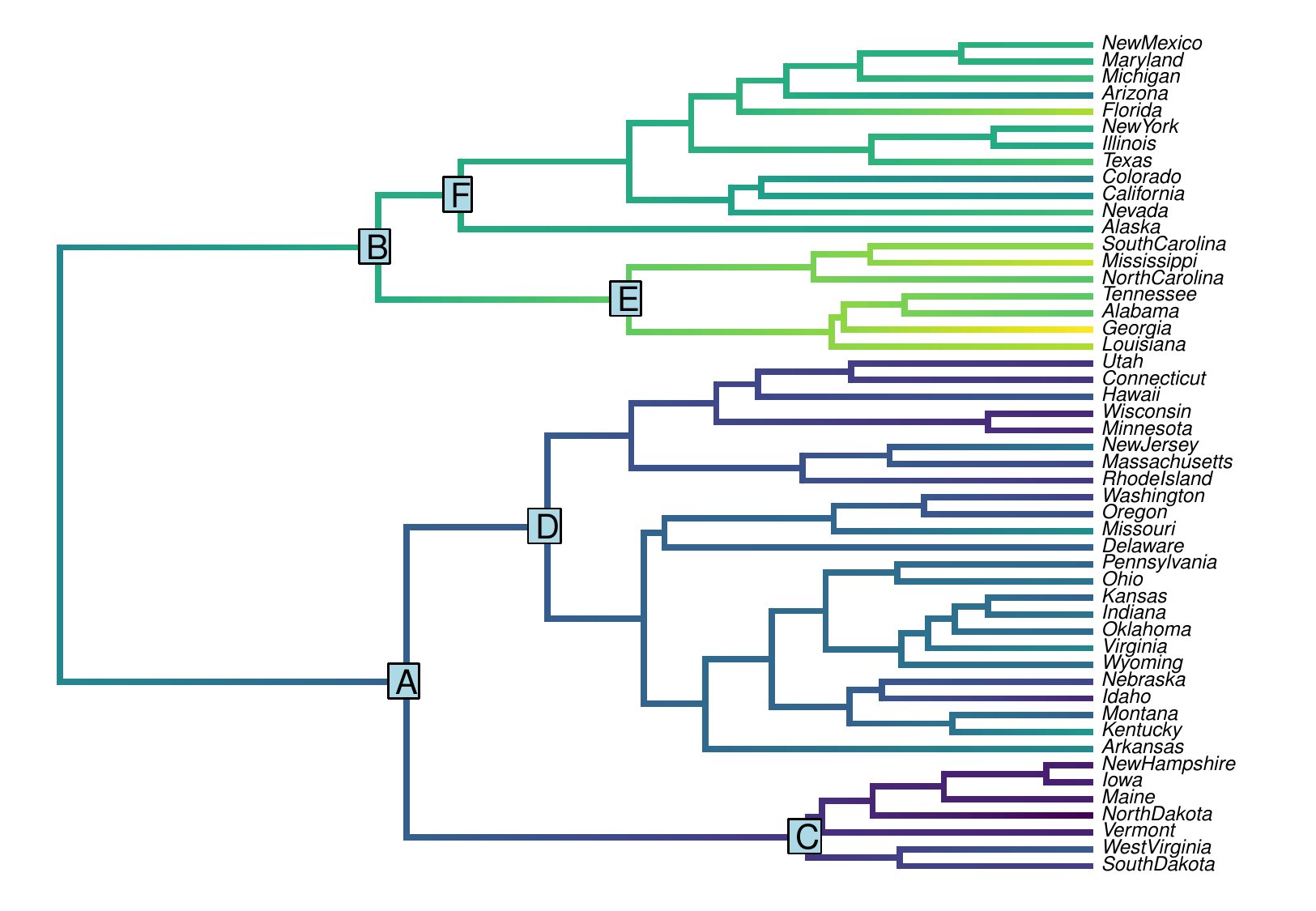}
  \captionof{figure}{Murder evolutionary dendrogram}
  \label{fig:USArrests_murder}
 \end{subfigure}
 \caption{Comparisson between the boxplot visualization and the evolutionary dendrogram on the murder feature from the USArrests dataset. \add{Evolutionary dendrograms provide more nuanced information as they take the full dendrogram hierarchy into account.}}
 \label{fig:boxplot_dendrogram_comparisson}
\end{figure}

Figure \ref{fig:boxplot_dendrogram_comparisson} 
compares such segmented boxplots to 
evolutionary dendrograms using 
the USArrests dataset\add{, which was analyzed in details (via evolutionary dendrograms) in Section \ref{sec:usarrests}}. First,
a dendrogram is obtained using 
the McQuitty method.
Next, segmented boxplots and 
the evolutionary dendrogram are
obtained for the \add{\textbf{Murder}} rate feature.

Figure \ref{fig:boxplots_usarrests} presents
boxplots grouped by cluster when
$2$, $3$ and $4$ clusters are used.
Using these plots, one can see that
the feature is well segmented for
the chosen number of clusters.
However, although boxplots bring 
measures of centrality and dispersion, 
they lose information about
the internal structure of each cluster.
In particular, one cannot find in the boxplots
which instances belong to each cluster.

In contrast Figure \ref{fig:USArrests_murder}
displays the evolutionary dendrogram obtained
for the same feature. By cutting the dendrogram
at the root, one obtains
the main top (B: green) and 
bottom (A: light blue) clusters.
By looking at the color dispersion in both clusters,
it is possible to conclude that
the top cluster has a larger variance than
the bottom cluster.
Furthermore, by looking at the next internal nodes,
one can observe that while the top cluster
can be further segmented into
dark green (F) and light green (E) sub-clusters, 
the bottom cluster can be segmented into
purple (C) and dark blue (D) sub-clusters.

\subsection{PFIS vs. \add{state-of-the-art} importance scores}
\label{sec:pfisRF}

This subsection compares 
PFIS, presented in 
subsection \ref{sec:importance}, to 
  common alternatives for \add{assessing}
feature importance in a dendrogram.
\add{We use the following baselines:
\begin{itemize}
    \item{[Classification-based importance - RF \citep{badih2019,ismaili2014}].} We first cut the dendrogram into $k$ clusters.
Next, these clusters are used as labels for
a random forest classifier \citep{breiman2001random} that uses the original features as inputs. Finally, the importance scores infered by
the random forest are used as
the importance of
the features for the dendrogram.
\item{[Classification-based importance - XGBoost \citep{badih2019,ismaili2014}].}  Same as before, but using a XGboost classifier \citep{chen2016xgboost} instead. 
\item{[LOVO \citep{badih2019}].} A leave-one-variable-out (LOVO) approach that compares within–cluster heterogeneity  by ommiting each feature from the dataset. This approach also requires the dendrogram to be cut into $k$ clusters.
\end{itemize}}

\add{In order to compare the performance of these approaches to PFIS, we need a dataset in which the ground-truth importance is know. Thus, we simulated data in a way that allows the true importance to be evaluated. Specifically,}
 let $X_{i,j}$ be
the $j$-th simulated feature value of 
the $i$-th instance, where
$1 \leq j \leq 6$ and $1 \leq i \leq 2^8$.
The data, $X$, is generated recursively 
according to a tree procedure described in
Algorithm \ref{alg:phylo_data}.
In this algorithm, $\theta_{i,j,k-1}$ represents
an ancestor of $\theta_{i,j,k}$ and
of $\theta_{2^{k-1}+i,j,k}$.
Also, $z_{i,j,k}$ represents the noise that
is added when generating a descendant. \add{Algorithm \ref{alg:feature_importance_simulated} summarizes this procedure.}
Observe that $z_{i,j,k}$ is 
of the order of magnitude of $(\sigma_j)^k$.
That is, the larger the value of $\sigma_j$,
the more noise is added to 
the leaf descendants proportionally to
the internal nodes of the tree.
As a result, the larger the value of $\sigma_j$,
the less the structure of the tree is
preserved by feature $j$.
Since in the simulation
$\sigma_j = 2^{j-3}$, one expects that
the feature importance in the adjusted dendrogram
should decrease from $j=1$ to $6$.

\begin{algorithm}
 \caption{ \small Simulated data \add{ to compare feature importance methods}}
 \label{alg:phylo_data}
 \algorithmicensure \ {\small 
 $X_{i,j}$, the simulated data.} \\
 \begin{algorithmic}[1]
  \State Let $\sigma = (2^{-2}, 2^{-1}, 2^0, 2^1, 2^2, 2^3)$.
  \State Let $\theta_{1,j,0} = 0$, for $1 \leq j \leq 6$.
  \For{each $k \in \{1,\ldots,7\}$,}
   \For{each $i \in \{1,\ldots,2^{k-1}\}$ and
   $j \in \{1 \ldots, 6\}$,}
   \State Let $i_1 = i$ and $i_2 = 2^{k-1}+i$
    \State Let $\theta_{i_1,j,k} 
    = \theta_{i_1,j,k-1} + z_{i_1,j,k}$,
    where $z_{i_1,j,k} \sim N(0, (\sigma_j)^k)$
    \State Let $\theta_{i_2,j,k} 
    = \theta_{i_1,j,k-1} + z_{i_2,j,k}$,
    where $z_{i_1,j,k} \sim N(0, (\sigma_j)^k)$
   \EndFor
  \EndFor
  \State Let $X_{i,j} = \theta_{i,j,7}$,
  for $1 \leq i \leq 2^8$ and $1 \leq j \leq 6$.
 \end{algorithmic}
 \label{alg:feature_importance_simulated}
\end{algorithm}

\begin{figure}[!hppt]
 \centering
 \includegraphics[scale=0.5]
 {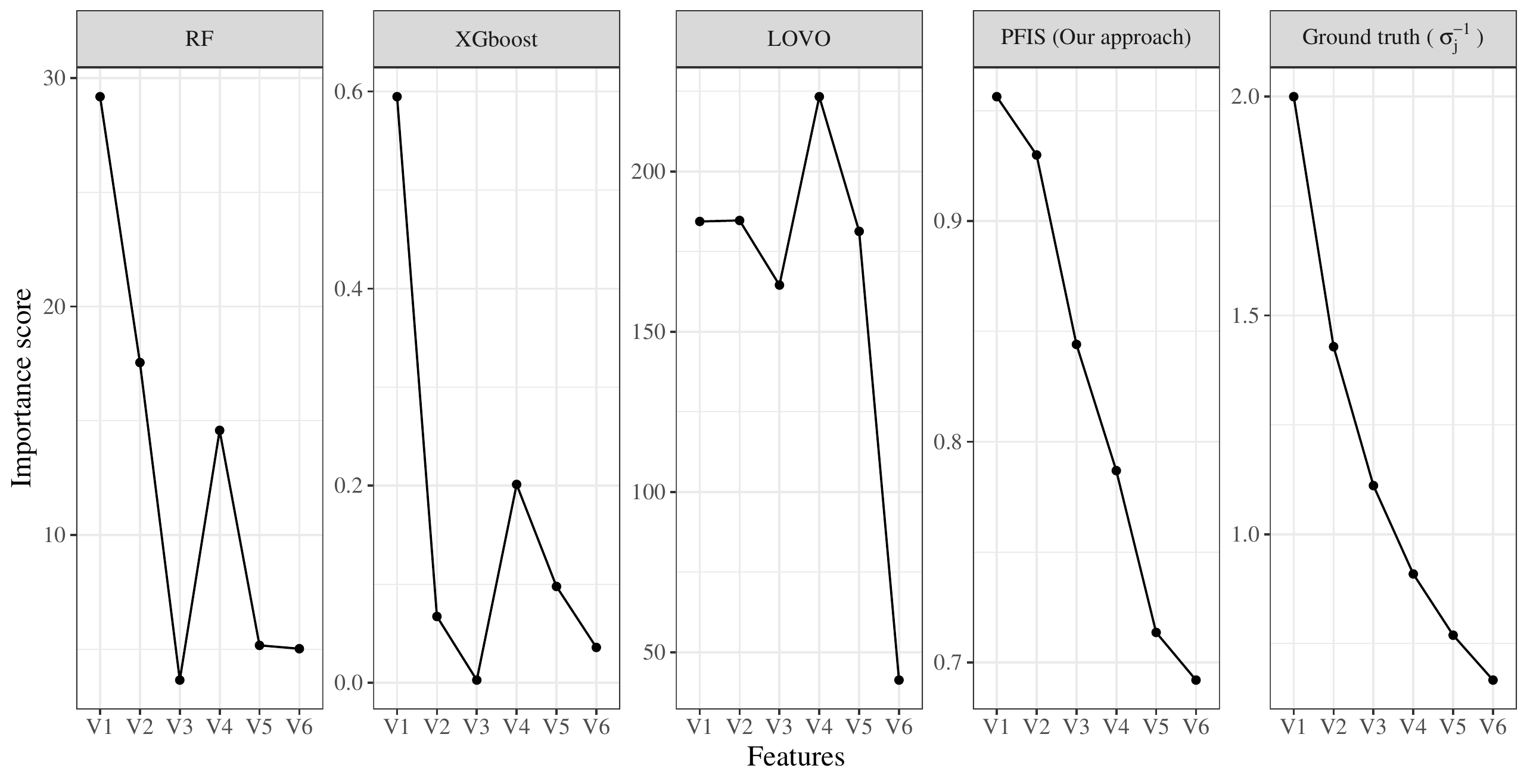}
 \caption{Measures of feature importance: \add{(far left) random forest \citep{ismaili2014}, (middle left) XGboost \citep{ismaili2014}, (middle) leave-one-variable-out \citep{badih2019}, (middle right) PFIS, (far right)  $\sigma_j^{-1}$}, 
 a gold standard \add{(ground truth)} that
 measures how much the tree structure that 
 is used to generate the data
 is preserved by each feature.  Our score preserves
the ordering  given by $\sigma_j^{-1}$, while   \add{all the others} do not. }
 \label{fig:importance_score_simulated}
\end{figure}

Figure \ref{fig:importance_score_simulated} shows \add{the various importance scores to}
 the data described above.
\add{The} dendrogram was obtained through
Ward's method\add{, and the scores that required a fixed number of clusters used $k = 3$ groups, which
  is the optimal number of clusters according to
the NbClust majority rule \citep{charrad2014nbclust}.}  
The right panel presents $\sigma_j^{-1}$,
a measure of how much the tree structure
is preserved by each feature. \add{The figure shows that our method (PFIS) is the only one to  preserve
the ordering of the features given by $\sigma_j^{-1}$, and thus uncover the true feature importance values.}

\subsection{Cross-validated Loss vs. \add{state-of-the-art dendrogram selection tools}} 
\label{sec:testsCVL}

\add{We use simulated and real data to compare our CVL loss to other state-of-the-art dendrogram selection tools.}
For each experiment we will use different combinations of clustering methods and linkages in order to assess the quality and performance of each methodology. The combinations of methods and linkages are listed \add{in Table \ref{tab:clust_comp}}.

This section evaluates the performance of 
cross-validated loss ($CVL$) by 
comparing it \add{to the following baselines:
\begin{itemize}
    \item {[Cophenetic Correlation score - COPH \citep{sokal1962comparison, timofeeva2019evaluating, saraccli2013comparison}]}. The idea of the cophenetic correlation is to compute the pearson correlation between the euclidean distance and the dendrogrammatic (or cophenetic) distance obtained for each different pair of observations. The dendrogrammatic distance is given by the height of the node of the obtained dendrogram at which each pair of observations are first joined together.
    \item {[Figure of Merit - FOM \citep{Yeung2001}]}. First, for each feature, a dendrogram is adjusted holding out the feature from the data. Next, by cutting each dendrogram at a given height, it is transformed into a partition. Finally, the within-cluster similarity is computed for each held-out feature using its respective partition. The FOM score corresponds to the average of these similarities.
    \item {[Adjusted Rand Index - ARI \citep{Hubert1985}]}. The standard rand index \citep{rand1971objective} measures the similarity between a  clustering of the data and a ground truth label by considering all pairs of observations and counting the pairs that are assigned to the same  cluster. ARI adds a correction to the rand index that takes into account assignment to the same clustering by chance.
\end{itemize}
}

\begin{table}[!hppt]
 \centering
 \caption{Clustering methods used 
 for comparing CVL to FOM\add{, COPH and ARI.}}
 \label{tab:clust_comp}
 \begin{tabular}{l|l}
  \textbf{Method} & \textbf{Linkages} \\
  \hline 
  Agnes 
  & weighted, average and ward \\
  Agglomerative 
  & ward.D, complete, single, ward.D2, \\
  & average, mcquitty and median \\
  Diana
  & Not applicable
 \end{tabular}
\end{table}

We compare CVL  \add{to these scores} using
simulated datasets and
UCI datasets\footnote{\url{https://archive.ics.uci.edu/ml/datasets.php}} that are available on $R$.
The chosen UCI datasets are summarized 
in \autoref{tab:datasets}.
We also use a simulated dataset with 
$250$ instances such that
the label is $Y_{i} \sim \text{Bernoulli}(0.5)$ and
the features are 
$(X_1,X_2)|Y \sim 
\text{N}_2(\begin{bmatrix} 2Y & 2Y\end{bmatrix}, \boldsymbol{I})$.
Note that although clustering methods
use only features, 
each dataset  also has a label. 

\begin{table}[h]
 \centering
 \caption{Datasets from UCI used to compare CVL to FOM.}
 \label{tab:datasets}
 \begin{tabular}{|c|c|c|c|}
  \hline
  \textbf{Dataset} 
  & \textbf{Sample size} 
  & \textbf{Number of features} 
  & \textbf{Label} \\ \hline
  Simulated & 250 & 2 & $Y$ \\ \hline
  Iris & 150 & 4 & Species \\ \hline
  Diabetes & 768 & 8 & $V9$ \\ \hline
  Wheat seeds & 210 & 7 & $V8$ \\ \hline
  Ionosphere & 351 & 34 & label \\ \hline
  Glass & 214 & 10 & $V11$ \\ \hline
  Haberman & 306 & 3 & $V4$ \\ \hline
  Wine & 178 & 13 & $V1$ \\ \hline
\end{tabular}
\end{table}

\add{We compare  all} methods to 
a gold-standard which 
uses the labels in the datasets.
The main idea is that
the dendrogram is treated as
a classifier for the label and
the F1 score is the gold-standard for
the dendrogram's performance.
Such a classifier is obtained by cutting
the dendrogram at an appropriate height
so that it becomes a partition with
size equal to the number of labels.
Next, each cluster in the partition is 
associated to a single label that
maximizes accuracy.
Each instance is labeled equally as
the partition to which it belongs.
The gold-standard for
the dendrogram's performance is
the F1 score based on its assigned labels.

Based on the above information,
it is possible to describe the experiment that
compares CVL, FOM, \add{Cophenetic correlation and ARI}.
For each dataset,
we performed clustering using
all of the the methods 
in \autoref{tab:clust_comp}.
Each of the $11$ clustering methods was applied using
the euclidean, manhattan and canberra distances.
As a result, $33$ dendrograms were obtained for each dataset.
For each dendrogram, the CVL, FOM, COPH, ARI and 
gold-standard F1 scores were calculated.
Since FOM \add{and ARI} require a pre-defined number of clusters,
we calculated \add{both} scores using the correct number of labels.
Since the correct number of clusters is
generally unavailable, this procedure gives
an advantage to FOM \add{and ARI}.

\autoref{tab:comps_HCPS_vs_FOM} shows,
for each dataset, the Spearman correlation between
$CVL$ and $F1$, $FOM$ and $F1$, \add{$-COPH$ and $F1$ and between $-ARI$ and $F1$} among
the clustering methods that were used.
Recall that, while CVL and FOM measure 
how bad is a given clustering,
\add{the gold standard} $F1$ \add{and the scores $ARI$ and $COPH$}  \add{are methods that} measure whether the clustering is good.
Therefore, a negative correlation between
a loss \add{or minus a score} and F1 indicates that
the measures are in agreement.
One can observe in \autoref{tab:comps_HCPS_vs_FOM} that,
while the Spearman correlation between CVL and F1 \add{and -COPH and F1} is
negative for all but one dataset, 
it is positive between FOM and F1 \add{and $-ARI$ and F1}
except for two \add{and five} datasets \add{respectively}.  
\add{Furthermore, both CVL and COPH outperformed the other methods in three out of the eight datasets each, while ARI performed better for two datasets.  Thus, these results provide evidence that both the cophenetic correlation and CVL  adequately assess 
the performance of clustering methods without the need to partition the dendrogram. In practice, they can complement each other   to perform dendrogram selection. } 

\begin{table}[H]
 \centering
 \caption{For each dataset, Spearman correlation
 between CVL and F1, CVL and FOM \add{, COPH and F1 and ARI and F1} among
 the $33$ clustering methods used in the experiment. \add{In order to maintain the comparison sign for all losses/scores, we compared minus the value of COPH and ARI to F1, as indicated by the sign in front of both scores}.
 Bold values indicate which of the losses\add{/scores} 
 has the lowest correlation with $F_1$ and, therefore,
 is in better agreement with this performance measure.}
 \label{tab:comps_HCPS_vs_FOM}
 \begin{tabular}{|c|c|c|c|c|}
  \hline
  \textbf{}
  & \multicolumn{4}{c|}{\textbf{Correlations}} \\ \hline
  \textbf{Dataset} & $CVL(\T)$ vs $F_1$ & $FOM$ vs $F_1$ &  $-COPH$ vs $F_1$ & $-ARI vs F_1$ \\ \hline
  Simulated        & \textbf{-0.40}   & 0.23      & -0.23  & 0.88      \\ \hline
  Iris             & 0.26             & 0.10 &  \textbf{-0.54} & -0.30 \\ \hline
  Pima indians     & -0.53   & 0.52      & \textbf{-0.63} & -0.06    \\ \hline
  Wheat seeds      & -0.11            & -0.48 &  0.28 &  \textbf{-0.83} \\ \hline
  Ionosphere       & \textbf{-0.43}   & 0.63 &  -0.12 & 0.60     \\ \hline
  Glass            & -0.15     & 0.19 &  \textbf{-0.48} & 0.14  \\ \hline
  Haberman         & \textbf{-0.42}   & 0.38   & -0.16    & -0.09       \\ \hline
  Wine             & -0.21             & -0.84 &  -0.35 & \textbf{-0.92} \\ \hline
 \end{tabular}
\end{table}

\section*{Acknowledgements}

L. M. C. C. is grateful for 
the scholarship provided by 
Funda\c{c}\~{a}o de Amparo \`{a} Pesquisa do 
Estado de S\~{a}o Paulo (FAPESP), grant 2020/10861-7.
R. I. is grateful for 
the financial support of FAPESP (grant 2019/11321-9) and 
CNPq (grant 309607/2020-5).
R. B. S. produced this work as
part of the activities of FAPESP Research, Innovation 
and Dissemination Center for Neuromathematics
(grant 2013/07699-0). 
The authors are also grateful for the suggestions given by 
Leonardo M. Borges and Victor Coscrato.

\bibliography{main}

\begin{thebibliography}{}

\bibitem[Badih et~al., 2019]{badih2019}
Badih, G., Pierre, M., and Laurent, B. (2019).
\newblock Assessing variable importance in clustering: a new method based on
  unsupervised binary decision trees.
\newblock {\em Computational Statistics}, 34:301--321.

\bibitem[Bollback, 2006]{bollback2006simmap}
Bollback, J.~P. (2006).
\newblock {SIMMAP}: stochastic character mapping of discrete traits on
  phylogenies.
\newblock {\em BMC bioinformatics}, 7(1):88.

\bibitem[Borges et~al., 2021]{borges2021}
Borges, L.~M., Izbicki, R., and Stern, R.~B. (2021).
\newblock The overlooked role of predictiveness in phylogenetics:
  Data-splitting as a powerful method for model selection.
\newblock {\em submitted}.

\bibitem[Breiman, 2001]{breiman2001random}
Breiman, L. (2001).
\newblock Random forests.
\newblock {\em Machine learning}, 45(1):5--32.

\bibitem[Caliński and Harabasz, 1974]{Calinski1974}
Caliński, T. and Harabasz, J. (1974).
\newblock A dendrite method for cluster analysis.
\newblock {\em Communications in Statistics-theory and Methods}, 3:1--27.

\bibitem[Charrad et~al., 2014]{charrad2014nbclust}
Charrad, M., Ghazzali, N., Boiteau, V., and Niknafs, A. (2014).
\newblock Nbclust: an r package for determining the relevant number of clusters
  in a data set.
\newblock {\em Journal of statistical software}, 61:1--36.

\bibitem[Chen and Guestrin, 2016]{chen2016xgboost}
Chen, T. and Guestrin, C. (2016).
\newblock Xgboost: A scalable tree boosting system.
\newblock In {\em Proceedings of the 22nd acm sigkdd international conference
  on knowledge discovery and data mining}, pages 785--794.

\bibitem[Chen et~al., 2014]{Chen2014}
Chen, Y., Kim, J., and Mahmassani, H. (2014).
\newblock Pattern recognition using clustering algorithm for scenario
  definition in traffic simulation-based decision support systems.

\bibitem[Clavel et~al., 2015]{Clavel2015}
Clavel, J., Escarguel, G., and Merceron, G. (2015).
\newblock mvmorph: an r package for fitting multivariate evolutionary models to
  morphometric data.
\newblock {\em Methods in Ecology and Evolution}, 6(11):1311--1319.

\bibitem[Coscrato et~al., 2019]{coscrato2019nls}
Coscrato, V., In{\'a}cio, M. H. d.~A., Botari, T., and Izbicki, R. (2019).
\newblock Nls: an accurate and yet easy-to-interpret regression method.
\newblock {\em arXiv preprint arXiv:1910.05206}.

\bibitem[Datta and Datta, 2003]{Datta2003}
Datta, S. and Datta, S. (2003).
\newblock {Comparisons and validation of statistical clustering techniques for
  microarray gene expression data}.
\newblock {\em Bioinformatics}, 19(4):459--466.

\bibitem[Felsenstein, 1985]{felsenstein1985phylogenies}
Felsenstein, J. (1985).
\newblock Phylogenies and the comparative method.
\newblock {\em The American Naturalist}, 125(1):1--15.

\bibitem[Fisher, 1936]{Fisher}
Fisher, R.~A. (1936).
\newblock The use of multiple measurements in taxonomic problems.
\newblock {\em Annals of eugenics}, 7(2):179--188.

\bibitem[{Fukunaga} and {Hostetler}, 1975]{Fukunaga1975}
{Fukunaga}, K. and {Hostetler}, L. (1975).
\newblock The estimation of the gradient of a density function, with
  applications in pattern recognition.
\newblock {\em IEEE Transactions on Information Theory}, 21(1):32--40.

\bibitem[Galili, 2015]{galili2015}
Galili, T. (2015).
\newblock {dendextend: an R package for visualizing, adjusting and comparing
  trees of hierarchical clustering}.
\newblock {\em Bioinformatics}, 31(22):3718--3720.

\bibitem[Hair, 2009]{hair2009multivariate}
Hair, J.~F. (2009).
\newblock Multivariate data analysis.

\bibitem[Hastie et~al., 2009a]{Elements2009}
Hastie, T., Tibshirani, R., and Friedman, J. (2009a).
\newblock {\em The Elements of Statistical Learning: Data Mining, Inference,
  and Prediction, Second Edition}.
\newblock Springer Series in Statistics. Springer New York.

\bibitem[Hastie et~al., 2009b]{hastie2009elements}
Hastie, T., Tibshirani, R., Friedman, J.~H., and Friedman, J.~H. (2009b).
\newblock {\em The elements of statistical learning: data mining, inference,
  and prediction}, volume~2.
\newblock Springer.

\bibitem[Hennig, 2015]{hennig2015clustering}
Hennig, C. (2015).
\newblock Clustering strategy and method selection.
\newblock {\em Handbook of cluster analysis}, 9:703--730.

\bibitem[Hubert and Arabie, 1985]{Hubert1985}
Hubert, L. and Arabie, P. (1985).
\newblock Comparing partitions.
\newblock {\em Journal of classification}, 2:193--218.

\bibitem[Ismaili et~al., 2014]{ismaili2014}
Ismaili, O.~A., Lemaire, V., and Cornu{\'e}jols, A. (2014).
\newblock A supervised methodology to measure the variables contribution to a
  clustering.
\newblock In Loo, C.~K., Yap, K.~S., Wong, K.~W., Teoh, A., and Huang, K.,
  editors, {\em Neural Information Processing}, pages 159--166, Cham. Springer
  International Publishing.

\bibitem[Izbicki and Lee, 2017]{izbicki2017converting}
Izbicki, R. and Lee, A.~B. (2017).
\newblock Converting high-dimensional regression to high-dimensional
  conditional density estimation.
\newblock {\em Electronic Journal of Statistics}, 11(2):2800--2831.

\bibitem[James et~al., 2013]{James2013introduction}
James, G., Witten, D., Hastie, T., and Tibshirani, R. (2013).
\newblock {\em An introduction to statistical learning}, volume 112.
\newblock Springer.

\bibitem[Jardine and {van Rijsbergen}, 1971]{jardine1971}
Jardine, N. and {van Rijsbergen}, C. (1971).
\newblock The use of hierarchic clustering in information retrieval.
\newblock {\em Information Storage and Retrieval}, 7(5):217 -- 240.

\bibitem[Joy et~al., 2016]{joy2016ancestral}
Joy, J.~B., Liang, R.~H., McCloskey, R.~M., Nguyen, T., and Poon, A.~F. (2016).
\newblock Ancestral reconstruction.
\newblock {\em PLoS computational biology}, 12(7):e1004763.

\bibitem[Kassambara, 2017]{kassambara2017}
Kassambara, A. (2017).
\newblock {\em Practical guide to cluster analysis in R: Unsupervised machine
  learning}, volume~1.
\newblock Sthda.

\bibitem[{Liu} et~al., 2006]{Liu2006}
{Liu}, J., {Bai}, Y., {Kang}, J., and {An}, N. (2006).
\newblock A new approach to hierarchical clustering using partial least
  squares.
\newblock In {\em 2006 International Conference on Machine Learning and
  Cybernetics}, pages 1125--1131.

\bibitem[MacQueen, 1967]{Macqueen1967}
MacQueen, J. (1967).
\newblock Some methods for classification and analysis of multivariate
  observations.
\newblock In {\em Proceedings of the Fifth Berkeley Symposium on Mathematical
  Statistics and Probability, Volume 1: Statistics}, pages 281--297, Berkeley,
  Calif. University of California Press.

\bibitem[McNeil, 1977]{USArrests}
McNeil, D.~R. (1977).
\newblock Interactive data analysis: a practical primer.

\bibitem[Neter et~al., 1996]{neter1996applied}
Neter, J., Kutner, M.~H., Nachtsheim, C.~J., Wasserman, W., et~al. (1996).
\newblock Applied linear statistical models.

\bibitem[O'Meara et~al., 2006]{OMeara2006}
O'Meara, B.~C., An{\'e}, C., Sanderson, M.~J., and Wainwright, P.~C. (2006).
\newblock Testing for different rates of continuous trait evolution using
  likelihood.
\newblock {\em Evolution}, 60(5):922--933.

\bibitem[Paradis and Schliep, 2019]{paradis2019ape}
Paradis, E. and Schliep, K. (2019).
\newblock ape 5.0: an environment for modern phylogenetics and evolutionary
  analyses in r.
\newblock {\em Bioinformatics}, 35(3):526--528.

\bibitem[Pfaffel, 2021]{FeatureImpCluster}
Pfaffel, O. (2021).
\newblock {\em FeatureImpCluster: Feature Importance for Partitional
  Clustering}.
\newblock R package version 0.1.5.

\bibitem[Pupko and Mayrose, 2020]{pupko2020gentle}
Pupko, T. and Mayrose, I. (2020).
\newblock A gentle introduction to probabilistic evolutionary models.

\bibitem[Rand, 1971]{rand1971objective}
Rand, W.~M. (1971).
\newblock Objective criteria for the evaluation of clustering methods.
\newblock {\em Journal of the American Statistical association},
  66(336):846--850.

\bibitem[Revell, 2012]{revell2012phytools}
Revell, L.~J. (2012).
\newblock phytools: an r package for phylogenetic comparative biology (and
  other things).
\newblock {\em Methods in ecology and evolution}, 3(2):217--223.

\bibitem[Revell, 2013]{revell2013two}
Revell, L.~J. (2013).
\newblock Two new graphical methods for mapping trait evolution on phylogenies.
\newblock {\em Methods in Ecology and Evolution}, 4(8):754--759.

\bibitem[Rocha et~al., 2009]{Rocha2009}
Rocha, L.~M., Cappabianco, F. A.~M., and Falcão, A.~X. (2009).
\newblock Data clustering as an optimum-path forest problem with applications
  in image analysis.
\newblock {\em International Journal of Imaging Systems and Technology},
  19(2):50--68.

\bibitem[Rosenberg and Hirschberg, 2007]{Andrew2007}
Rosenberg, A. and Hirschberg, J. (2007).
\newblock V-measure: A conditional entropy-based external cluster evaluation
  measure.
\newblock pages 410--420.

\bibitem[Rousseeuw, 1987]{Peter1987}
Rousseeuw, P.~J. (1987).
\newblock Silhouettes: a graphical aid to the interpretation and validation of
  cluster analysis.
\newblock {\em Computational and Applied Mathematics}, 20:53--65.

\bibitem[Rousseeuw and Kaufman, 1990]{Rousseeuw1990}
Rousseeuw, P.~J. and Kaufman, L. (1990).
\newblock Finding groups in data.
\newblock {\em Hoboken: Wiley Online Library}, 1.

\bibitem[Sara{\c{c}}li et~al., 2013]{saraccli2013comparison}
Sara{\c{c}}li, S., Do{\u{g}}an, N., and Do{\u{g}}an, {\.I}. (2013).
\newblock Comparison of hierarchical cluster analysis methods by cophenetic
  correlation.
\newblock {\em Journal of inequalities and Applications}, 2013(1):1--8.

\bibitem[Seo and Shneiderman, 2002]{jinwook2002}
Seo, J. and Shneiderman, B. (2002).
\newblock Interactively exploring hierarchical clustering results [gene
  identification].
\newblock {\em Computer}, 35(7):80--86.

\bibitem[Shimizu et~al., 2022]{shimizu2022model}
Shimizu, G.~Y., Izbicki, R., and de~Carvalho, A.~C. (2022).
\newblock Model interpretation using improved local regression with variable
  importance.
\newblock {\em arXiv preprint arXiv:2209.05371}.

\bibitem[Sokal and Rohlf, 1962]{sokal1962comparison}
Sokal, R.~R. and Rohlf, F.~J. (1962).
\newblock The comparison of dendrograms by objective methods.
\newblock {\em Taxon}, pages 33--40.

\bibitem[Tibshirani, 1996]{tibshirani1996regression}
Tibshirani, R. (1996).
\newblock Regression shrinkage and selection via the lasso.
\newblock {\em Journal of the Royal Statistical Society: Series B
  (Methodological)}, 58(1):267--288.

\bibitem[Timofeeva, 2019]{timofeeva2019evaluating}
Timofeeva, A. (2019).
\newblock Evaluating the robustness of goodness-of-fit measures for
  hierarchical clustering.
\newblock In {\em Journal of Physics: Conference Series}, volume 1145, page
  012049. IOP Publishing.

\bibitem[Van~der Maaten and Hinton, 2008]{van2008visualizing}
Van~der Maaten, L. and Hinton, G. (2008).
\newblock Visualizing data using t-sne.
\newblock {\em Journal of machine learning research}, 9(11).

\bibitem[Vendramin et~al., 2010]{vendramin2010relative}
Vendramin, L., Campello, R.~J., and Hruschka, E.~R. (2010).
\newblock Relative clustering validity criteria: A comparative overview.
\newblock {\em Statistical analysis and data mining: the ASA data science
  journal}, 3(4):209--235.

\bibitem[Ward~Jr, 1963]{ward1963hierarchical}
Ward~Jr, J.~H. (1963).
\newblock Hierarchical grouping to optimize an objective function.
\newblock {\em Journal of the American statistical association},
  58(301):236--244.

\bibitem[Yeung et~al., 2001]{Yeung2001}
Yeung, K.~Y., Haynor, D.~R., and Ruzzo, W.~L. (2001).
\newblock {Validating clustering for gene expression data}.
\newblock {\em Bioinformatics}, 17(4):309--318.

\bibitem[Zhu, 2018]{zhu2018variable}
Zhu, X. (2018).
\newblock Variable diagnostics in model-based clustering through variation
  partition.
\newblock {\em Journal of Applied Statistics}, 45(16):2888--2905.

\end{thebibliography}

\end{document}